\documentclass{aa}
\usepackage{txfonts}
\usepackage{graphicx}

\begin{document}

\title{Galaxy And Mass Assembly (GAMA): The environmental dependence of the galaxy main sequence}
\author{L. Wang\inst{1,2,3}, P. Norberg\inst{3}, S. Brough\inst{4}, M. J. I. Brown\inst{5}, E. da Cunha\inst{6}, L. J. Davies\inst{7},  S. P. Driver\inst{7}, B. W. Holwerda\inst{8}, A. M. Hopkins\inst{9}, M. A. Lara-Lopez\inst{10}, J. Liske\inst{11}, J. Loveday\inst{12}, M. W. Grootes\inst{13, 14}, C. C. Popescu\inst{15, 16}, A. H. Wright\inst{17}}

\institute{SRON Netherlands Institute for Space Research, Landleven 12, 9747 AD, Groningen, The Netherlands \email{l.wang@sron.nl} 
\and Kapteyn Astronomical Institute, University of Groningen, Postbus 800, 9700 AV Groningen, the Netherlands
\and ICC \& CEA, Department of Physics, Durham University, Durham, DH1 3LE, UK
\and School of Physics, University of New South Wales, NSW 2052, Australia
\and School of Physics and Astronomy, Monash University, Victoria 3800, Australia
\and Research School of Astronomy and Astrophysics, Australian National University, Canberra, ACT 2611, Australia
\and International Centre for Radio Astronomy Research (ICRAR), University of Western Australia, Crawley, WA 6009, Australia
\and Department of Physics and Astronomy, 102 Natural Science Building, University of Louisville, Louisville KY 40292, USA
\and Australian Astronomical Observatory, 105 Delhi Rd, North Ryde, NSW 2113, Australia
\and Dark Cosmology Centre, Niels Bohr Institute, University of Copenhagen, Juliane Maries Vej 30, DK-2100 Copenhagen, Denmark
\and Hamburger Sternwarte, Universit\"at Hamburg, Gojenbergsweg 112, 21029 Hamburg, Germany
\and Astronomy Centre, University of Sussex, Falmer, Brighton BN1 9QH, UK
\and Max-Planck-Institut f\"ur Kernphysik, Saupfercheckweg 1, 69117 Heidelberg, Germany
\and ESA/ESTEC SCI-S, Keplerlaan 1, 2201 AZ, Noordwijk, The Netherlands
\and Jeremiah Horrocks Institute, University of Central Lancashire, PR1 2HE, Preston, UK
\and The Astronomical Institute of the Romanian Academy, Str. Cutitul de Argint 5, Bucharest, Romania
\and Argelander-Institut f\"ur Astronomie, Universit\"at Bonn, Auf dem H\"gel 71, D-53121 Bonn, Germany}

\date{Received / Accepted}

\abstract
   {}
   {We aim to investigate if the environment (characterised by the host dark matter halo mass) plays any role in shaping the galaxy star formation main sequence (MS).}
   {The Galaxy and Mass Assembly project (GAMA) combines a spectroscopic survey with photometric information in 21 bands from the far-ultraviolet (FUV) to the far-infrared (FIR). Stellar masses and dust-corrected star-formation rates (SFR) are derived from spectral energy distribution (SED) modelling using MAGPHYS. We use the GAMA galaxy group catalogue to examine the variation of the fraction of star-forming galaxies (SFG) and properties of the MS with respect to the environment.}
   {We examine the environmental dependence for stellar mass selected samples without preselecting star-forming galaxies and study any dependence on the host halo mass separately for centrals and satellites  out to $z\sim0.3$. We find the SFR distribution at fixed stellar mass can be described by the combination of two Gaussians (referred to as the star-forming Gaussian and the quiescent Gaussian).  Using the observed bimodality to define SFG, we investigate how the fraction of SFG F(SFG) and properties of the MS change with environment. For centrals, the position of the MS is similar to the field but with a larger scatter. No significant dependence on halo mass is observed. For satellites, the position of the MS is almost always lower (by $\sim0.2$ dex) compared to the field and the width is almost always larger. F(SFG) is similar between centrals (in different halo mass bins) and field galaxies. However, for satellites F(SFG) decreases with increasing halo mass and this dependence is stronger towards lower redshift.}
   {}
   

\titlerunning{The environmental dependence of the galaxy main sequence}

\authorrunning{Wang et al.}

\maketitle

\section{Introduction}

Star-forming galaxies exhibit a tight correlation between star-formation rate (SFR) and stellar mass, known as the galaxy star formation main sequence (MS). The normalisation of the MS evolves with time such that galaxies at fixed stellar mass have increasingly higher SFR in the distant Universe (e.g.,  Daddi et al. 2007; Elbaz et al. 2007; Noeske et al. 2007; Wang et al. 2013; Schreiber et al. 2015). The discovery of the MS has contributed significantly in overturning our previous thinking which involved most galaxies undergoing chaotic processes to a dramatically different picture which implies star formation in the majority of galaxies is governed by quasi-steady processes. Detailed studies of the MS, e.g.,  how it depends on the environment, are crucial for our understanding of galaxy evolution. Peng et al. (2010) found that the MS (its slope and width) is independent of environment and only the fraction of star-forming galaxies at fixed stellar mass change with environment. Subsequently, many studies have attempted to identify  environmental dependence of the MS, but a consensus is still lacking. Lin et al. (2014) showed that MS is indistinguishable between the field and group environment out to $z\sim0.8$, but reported a moderate SFR decrease in clusters. Koyama et al. (2013) compiled an H$\alpha$-selected galaxy sample in clusters and demonstrated that any potential environmental impact is small ($\lesssim0.2$ dex) since $z\sim2$. Erfanianfar et al. (2015) concluded that the MS of group galaxies (in halos around $[10^{12.5}, 10^{14.2}]M_{\odot}$) deviates from the MS in the field but this environmental effect weakens with increasing redshift. Alpaslan et al. (2016) looked at the dependence of the MS of disks (from a morphologically selected sample) as a function of location with respect to the filamentary large-scale structure out to $z=0.13$ and found that at fixed stellar mass the SFRs of spiral galaxies in filaments are higher on the periphery of the filament compared to its core. Using the same sample of disk galaxies, Grootes et al. (2017) investigated the environment dependence of the MS for central and satellite spiral galaxies. They found that the MS for central spirals are similar to field spirals but the MS for satellite spirals exhibit a median offset ($\sim0.1-0.2$ dex) compared to the field population.

In this paper, we take advantage of the Galaxy and Mass Assembly survey (GAMA; Driver et al. 2011) galaxy group catalogues to examine the fraction of star-forming galaxies and the distribution of central and satellite galaxies in the SFR versus stellar mass parameter space as a function of environment (characterised by the host halo mass) and redshift. The paper is organised as follows. In Section 2, we briefly describe the GAMA spectroscopic survey and the associated multi-wavelength photometric data. We summarise the procedures used to derive stellar mass, SFR and halo mass of GAMA groups. The main results are presented in Section 3. Conclusions and discussions are given in Section 4. We assume $\Omega_m=0.3$,  $\Omega_{\Lambda}=0.7$,  and $H_0=70$ km s$^{-1}$ Mpc$^{-1}$. Flux densities are corrected for Galactic extinction (Schlegel, Finkbeiner \& Davis 1998).

\section{Data}

The GAMA survey has been operating since 2008 to acquire spectra for galaxies selected from the SDSS (York et al. 2000). Reliable redshifts have been obtained for over 98\% of GAMA galaxies ($\sim$200,000 in total) with r < 19.8 mag (Liske et al. 2015) which is 2 magnitudes deeper than the SDSS main  survey. Our target catalogue covers three roughly equal-sized ($12\times5$ deg$^2$ each) regions centred at a right ascension of 9h, 12h and 14.5h, on the celestial equator. We impose a lower redshift limit of 0.01 to remove objects for which peculiar motion could overwhelm the Hubble flow and an upper limit of 0.3 as very few galaxies are above this redshift. The GAMA team also assembled photometric information in 21 bands from the far-ultraviolet (FUV) to the far-infrared (FIR). Wright et al. (2016) presented the GAMA LAMBDAR Data Release (LDR) containing 21-band deblended matched aperture photometry using apertures defined in the SDSS r-band for all images in the imaging dataset. The LDR is specifically designed for spectral energy distribution (SED) modelling, as the photometry and uncertainties are consistently measured across the entire bandpass.

\subsection{Stellar mass and star-formate rate estimates}

We use stellar masses and dust-corrected SFRs derived by fitting panchromatic SEDs to the full 21-band dataset LDR using MAGPHYS (da Cunha et al. 2008; da Cunha \& Charlot 2011), assuming Bruzual \& Charlot (2003) (BC03) models, a Chabrier (2003) initial mass function, and the Charlot \& Fall (2000) dust attenuation law. A detailed description of the MAGPHYS fits to the LDR can be found in Driver et. al. (2016). Wright et al. (2017) found that the MAGPHYS stellar masses agree well with stellar masses derived in Taylor et al. (2011) (within 0.2 dex for 95\% of the sample). Similarly, reasonably good agreements (within 0.3 dex) are observed between MAGPHYS UV-IR SED fitting based SFRs and other SFR estimates (Davies et al. 2016; Driver et al. 2018). We divide our sample into three bins, $z1=[0.01, 0.1]$, $z2=[0.1, 0.2]$ and $z3=[0.2, 0.3]$, and use conservative stellar mass limit estimates, $\log M^{\rm limit}_{\rm star}=$9.49, 10.44, 10.82 at $z=0.1$, 0.2, 0.3 respectively, which are designed to be 100\% complete particularly with respect to bias in colour (Wright et al. 2017).

\begin{figure}
\centering
\includegraphics[height=2.in,width=3.in]{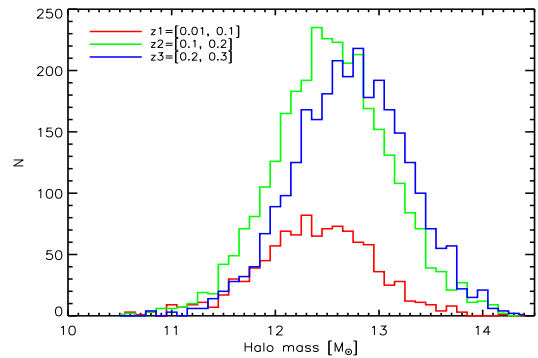}
\caption{The host dark matter halo mass distribution in three redshift bins. We focus on the halo mass range between $10^{12}$ and $10^{14} M_{\odot}$.}
\label{halomass}
\end{figure}

\subsection{Host dark matter halo mass estimates}

We use the classification of galaxies into groups as listed in the GAMA Galaxy Group Catalogue generated using a redshift space friends-of-friends group finding algorithm (Robotham et al. 2011). The algorithm has been tested extensively on mock GAMA lightcones (Merson et al. 2013). Galaxies are classified as either the central or a satellite of their group. Recovered group properties are robust to the effects of interlopers and are median unbiased in the most important respects (such as halo mass estimates). In total, the Group Catalogue contains over 23,000 galaxy groups (with a minimum multiplicity of 2 members\footnote{Our general conclusions are unaffected by changing the minimum multiplicity from 2 to 5.}). Host halo mass is derived using a variety of methods such as weak lensing calibrated dynamically or luminosity derived masses and abundance matching derived masses. We use weak lensing calibrated dynamically derived masses as our default, although we our results are robust against different choices of halo mass estimators. Robotham et al. (2011) found that the most accurate recovery of the dynamical centre of the group is obtained using the so-called iterative group centre which always coincides with a group member galaxy. We use this galaxy as the central galaxy of the group, and so all other member galaxies are treated as satellites. Fig.~\ref{halomass} shows the host  halo mass distribution of the GAMA groups in three redshift bins. Most groups have host halo masses between $10^{12}$ and $10^{14}$ $M_{\odot}$ and so this is the mass range we will focus on.

\section{Results}

\begin{figure}
\centering
\includegraphics[height=0.9in,width=1.15in]{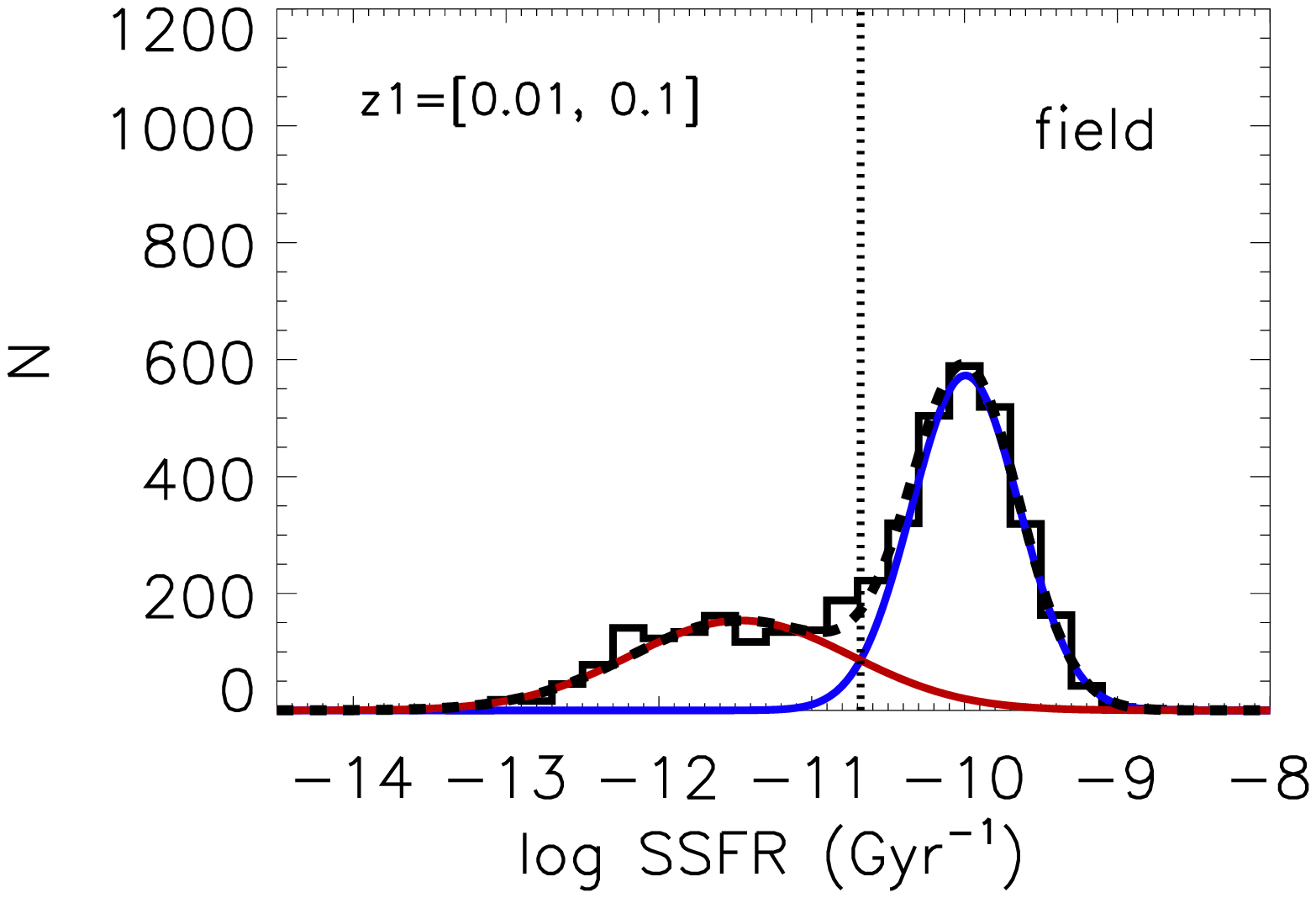}
\includegraphics[height=0.9in,width=1.15in]{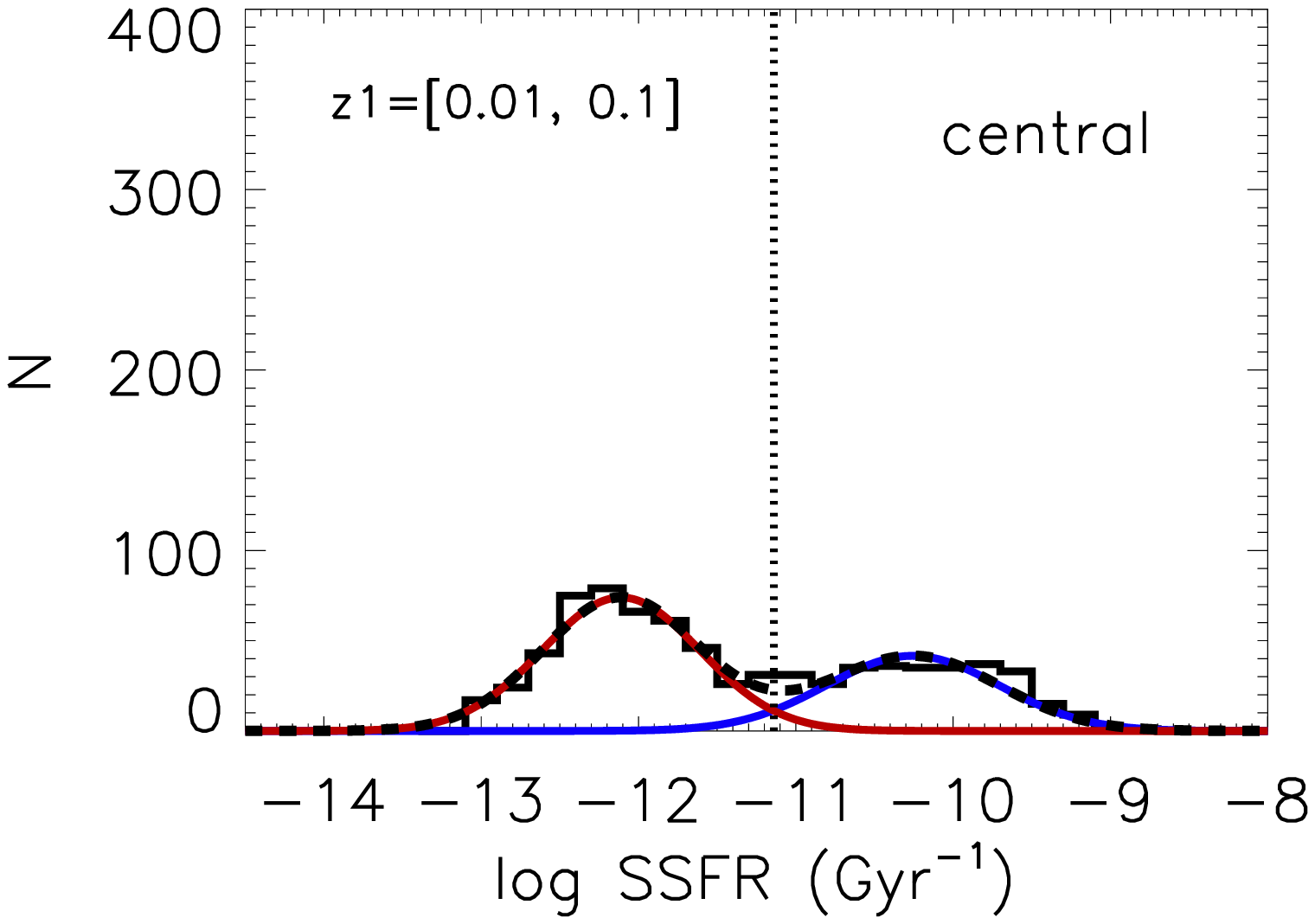}
\includegraphics[height=0.9in,width=1.15in]{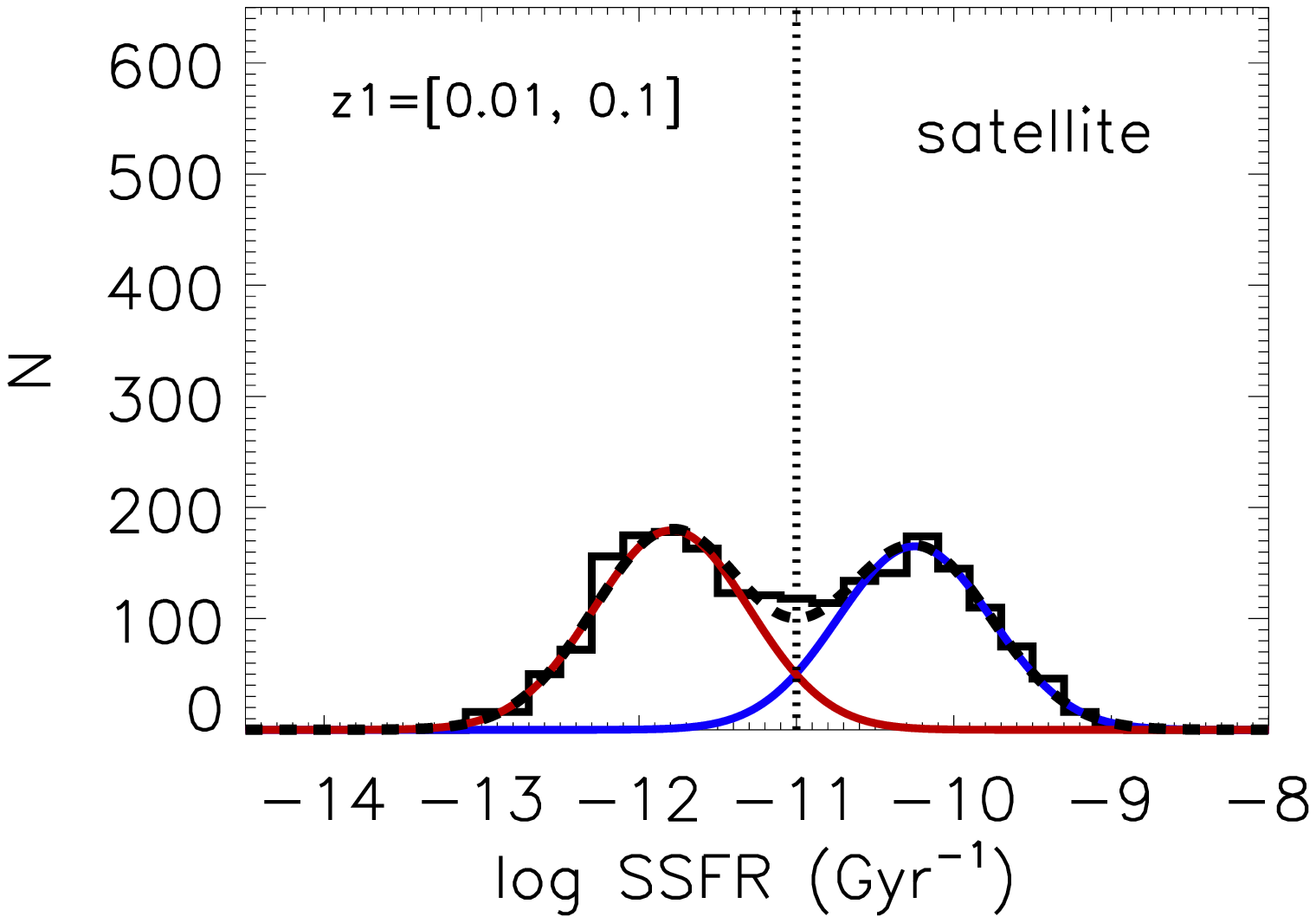}
\includegraphics[height=0.9in,width=1.15in]{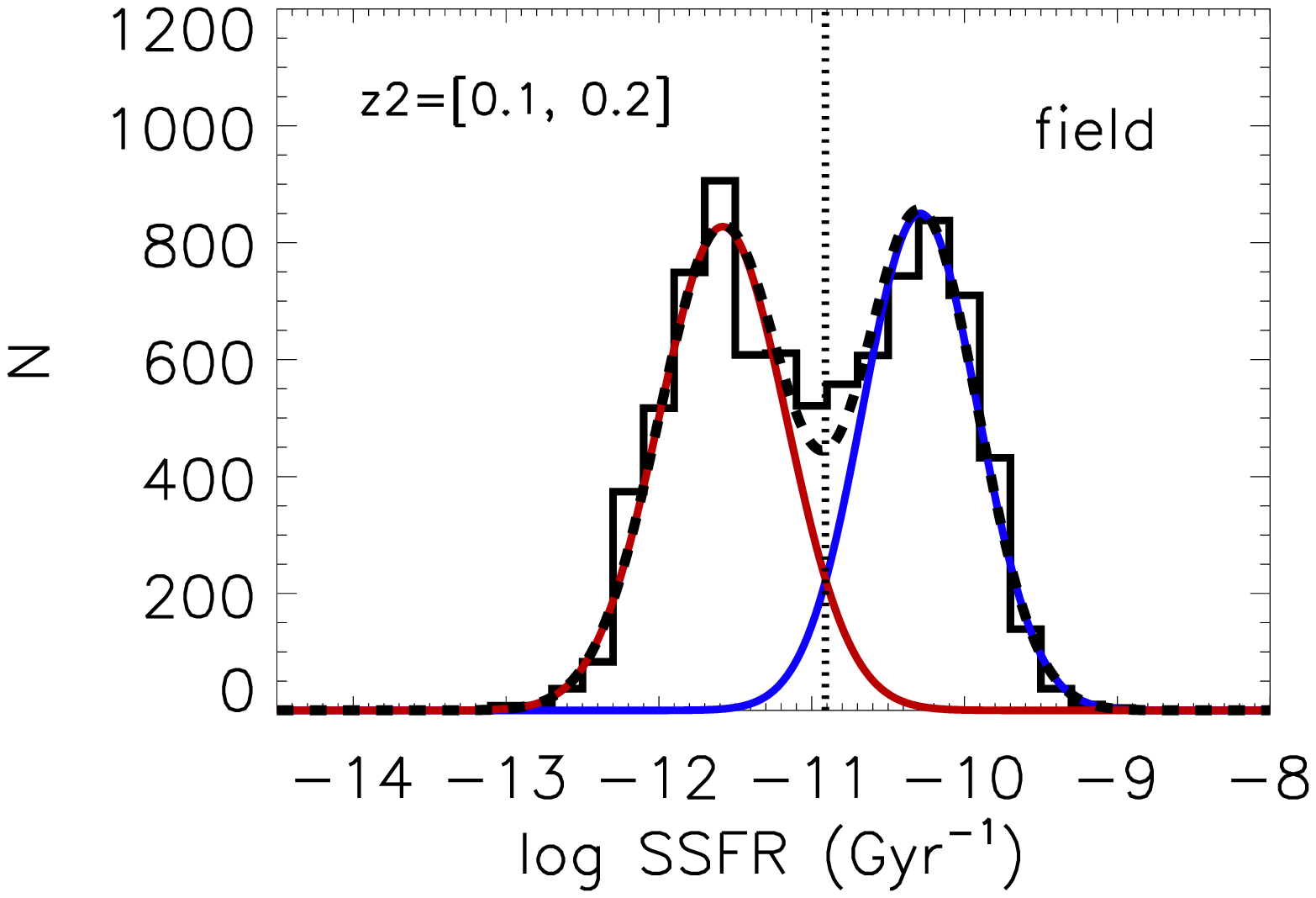}
\includegraphics[height=0.9in,width=1.15in]{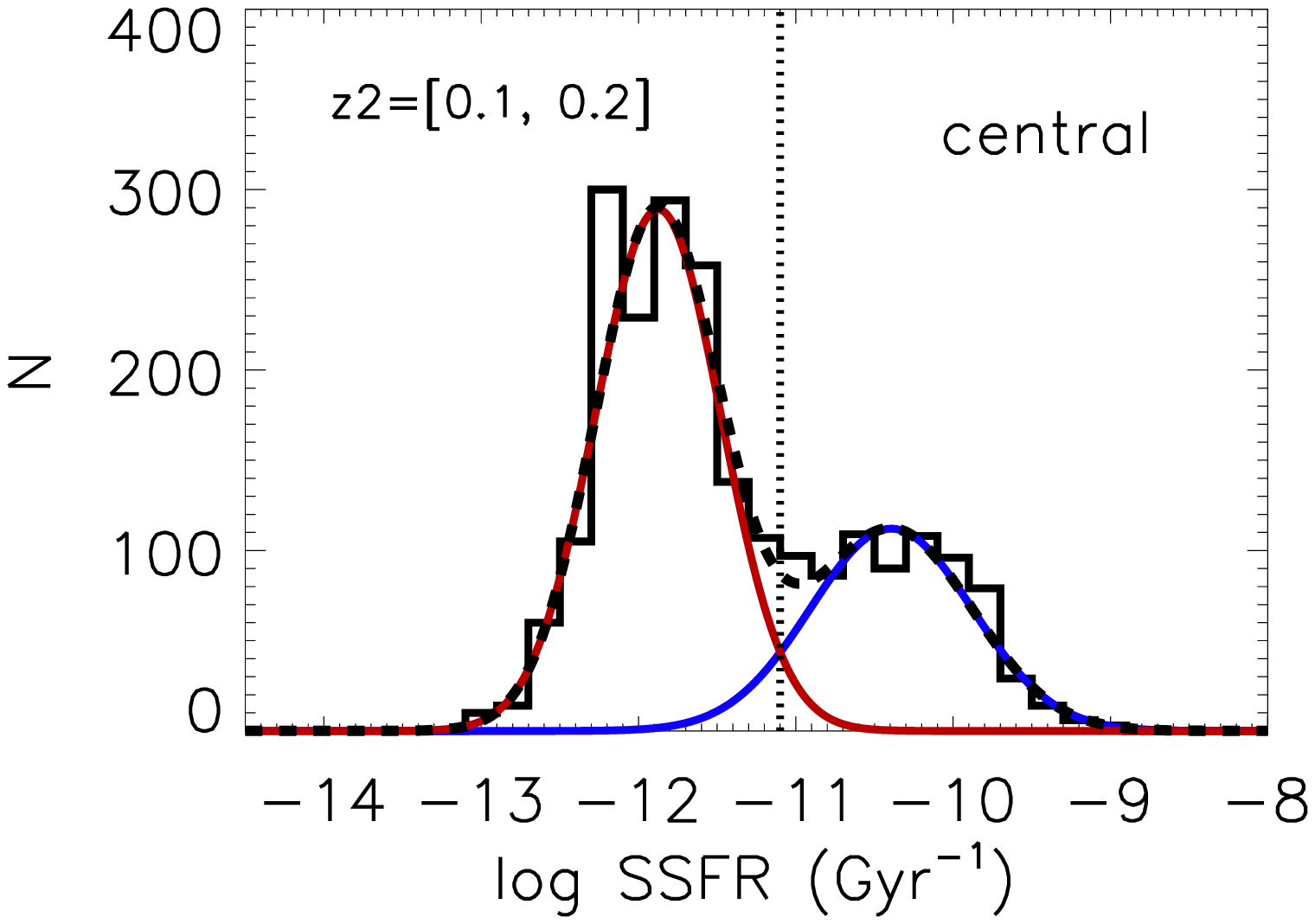}
\includegraphics[height=0.9in,width=1.15in]{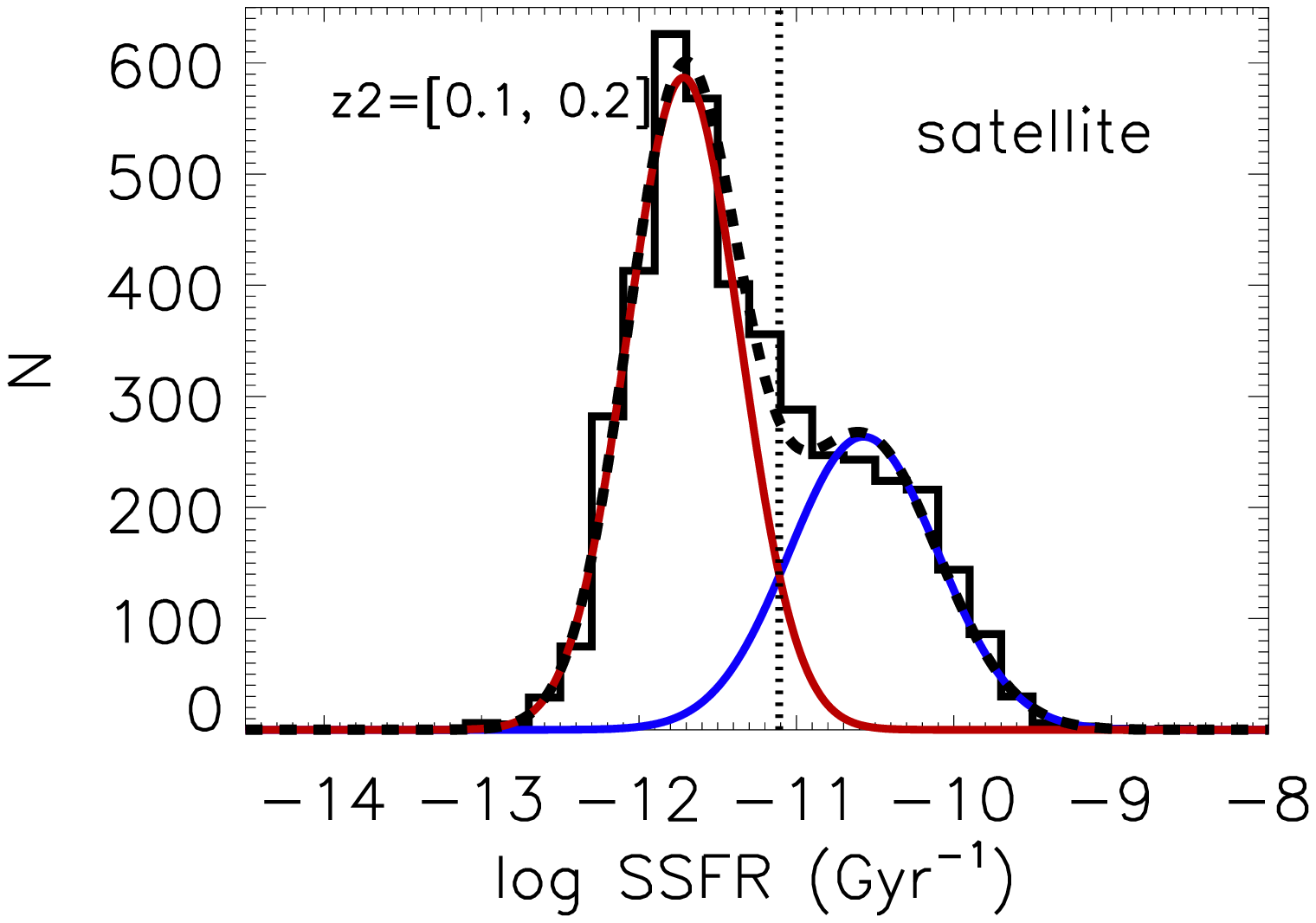}
\includegraphics[height=0.9in,width=1.15in]{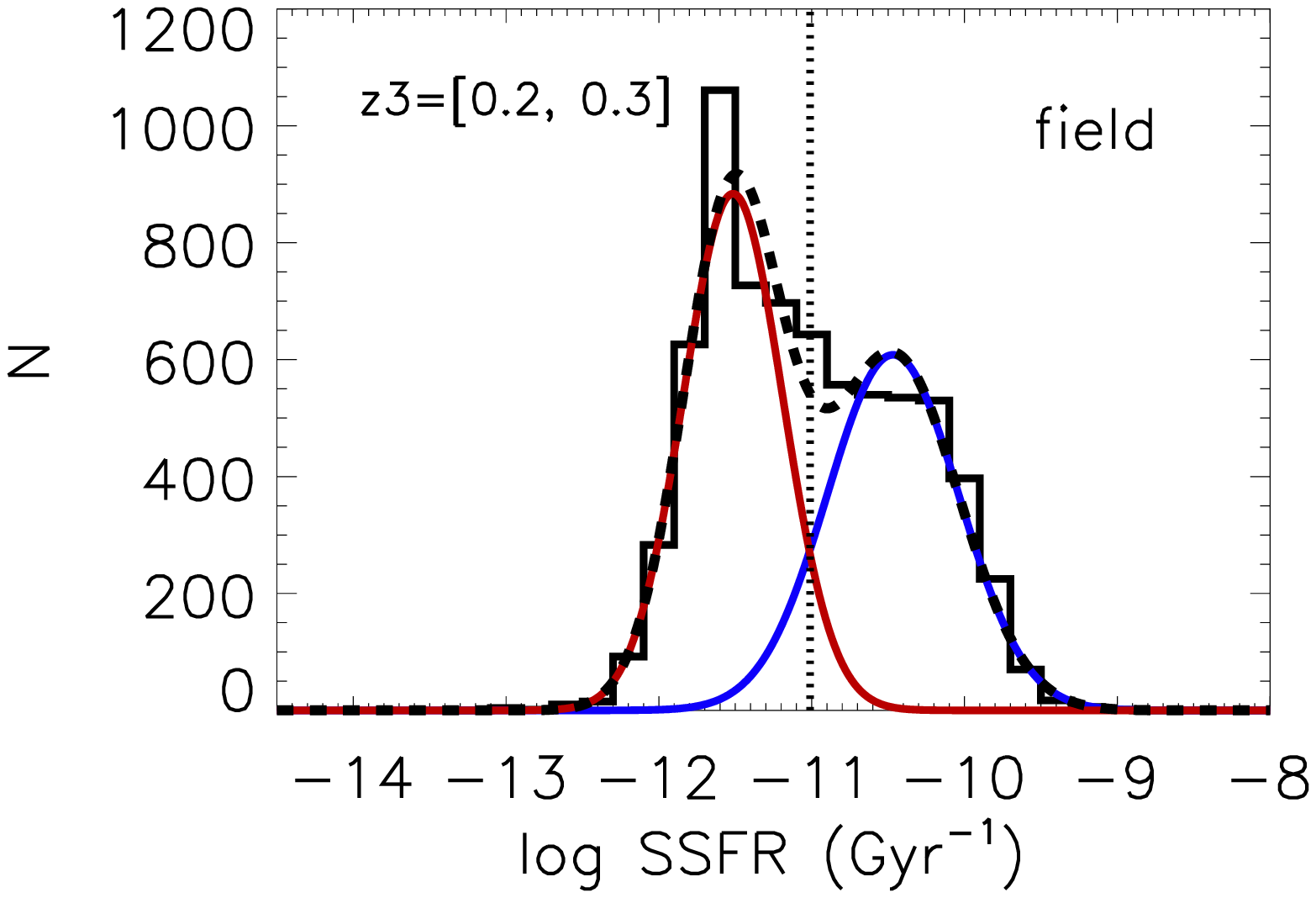}
\includegraphics[height=0.9in,width=1.15in]{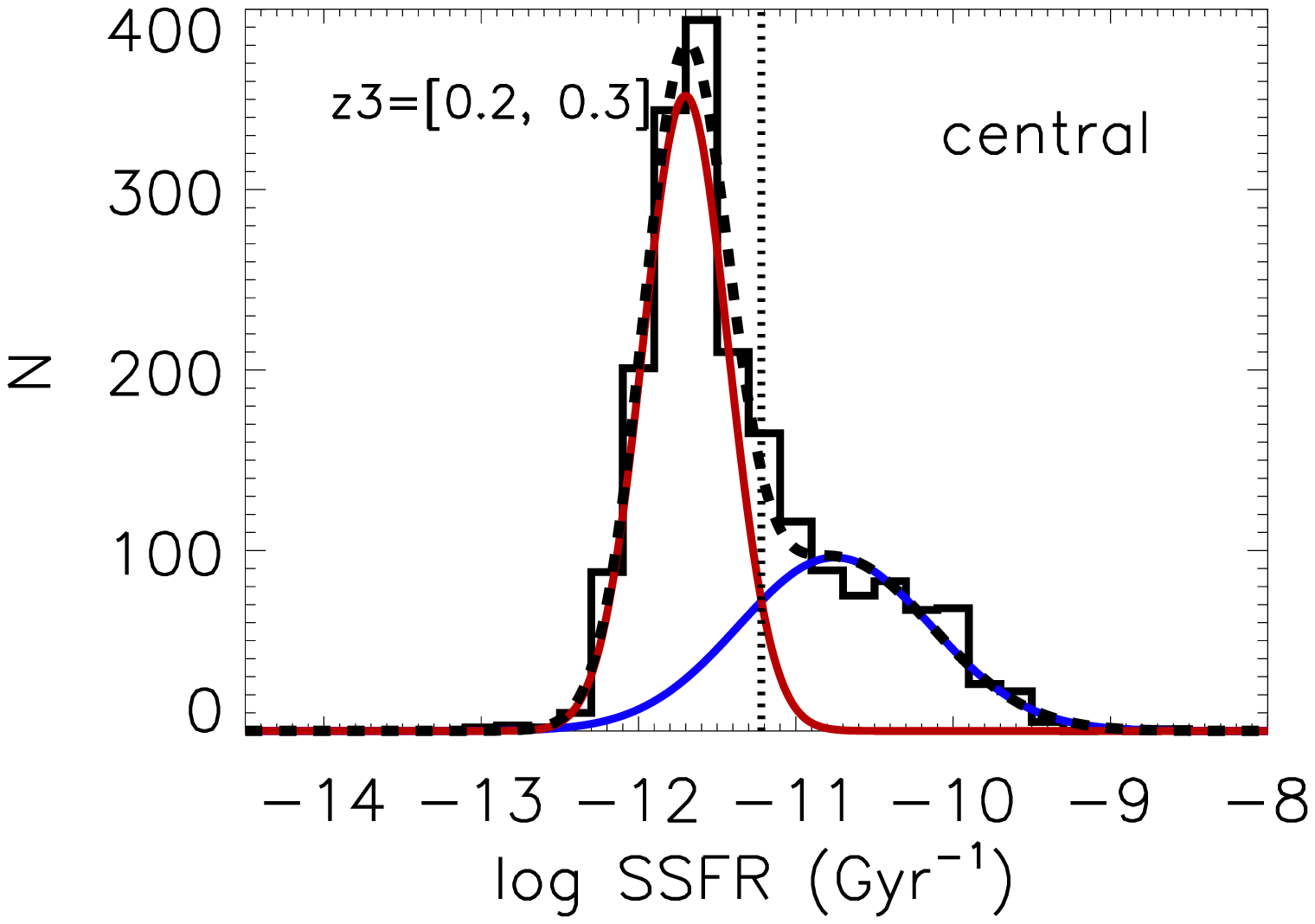}
\includegraphics[height=0.9in,width=1.15in]{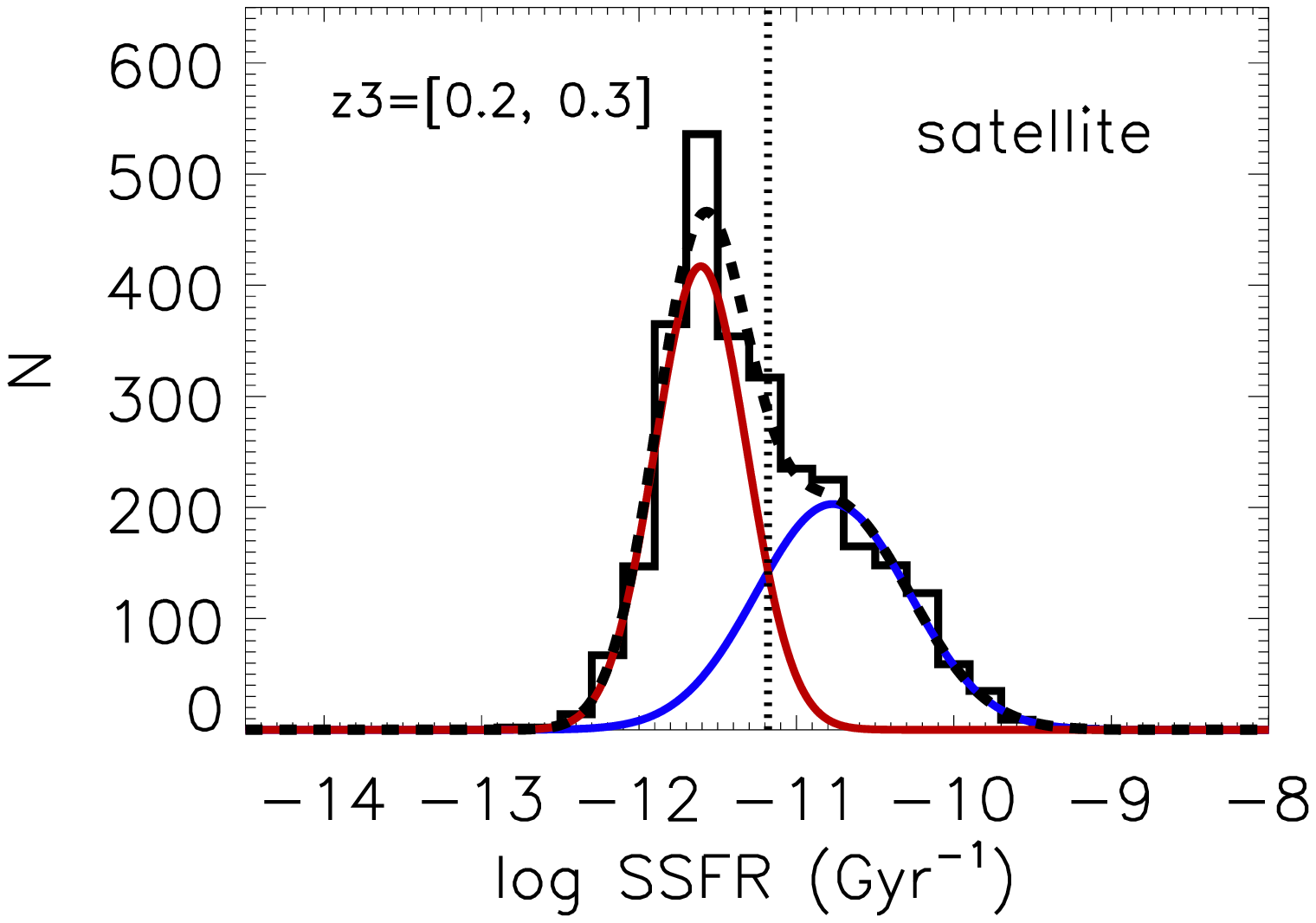}
\caption{The $\log$ SSFR distribution for field, central and satellite galaxies, which can be fit by the combination of two Gaussians (the blue line and the red line). The dashed line is the sum of the two. The dotted line indicates where the blue line is equal to the red line in amplitude.}
\label{ssfr}
\end{figure}

The MS  seems to be already in place at $z\sim4$ or even earlier (e.g., Dunne et al. 2009; Shim et al. 2011; Lee et al. 2012; Steinhardt 2014). However, important aspects of the MS which are still under-explored are any potential environmental dependence and how it might change with redshift.

\subsection{Overall analysis}

To study the environmental dependence of the MS, first we need to define what we mean by star-forming galaxies. In Fig.~\ref{ssfr}, we plot the distribution of the $\log$ of the specific SFR (SSFR) for field, central and satellite galaxies in three redshift bins. As described in Section 2.2, we define group galaxies to be those in groups with host halo mass between $10^{12}$ and $10^{14} M_{\odot}$. Consequently, galaxies not classified as group galaxies are called field galaxies. The group galaxies are further divided into centrals and satellites. The $\log$ SSFR distribution is clearly bimodal (consistent with the bimodality discovered in the colour-mass diagram in Taylor et al. 2015) and can be approximated by the combination of two Gaussians. The blue line (referred to hereafter as the "star-forming Gaussian") is the best-fit Gaussian to the galaxies with typically higher SSFR and the red line (referred to as the "quiescent Gaussian") is the best-fit to galaxies with typically lower SSFR. The dashed line is the sum of the two Gaussians. The dotted line indicates where the amplitude of the blue line is equal to that of the red line (referred to as the threshold $T$). We list best-fit values of the location and scale parameters of the two Gaussians, the threshold log SSFR value ($T$) and the number of galaxies for each panel in Fig. 2 in Table A.1 in the Appendix. Note that the threshold $T$ values are very close to -11 (Gyr$^{-1}$) in different redshift bins and for different galaxy types.

\begin{figure}
\centering
\includegraphics[height=1.3in,width=1.75in]{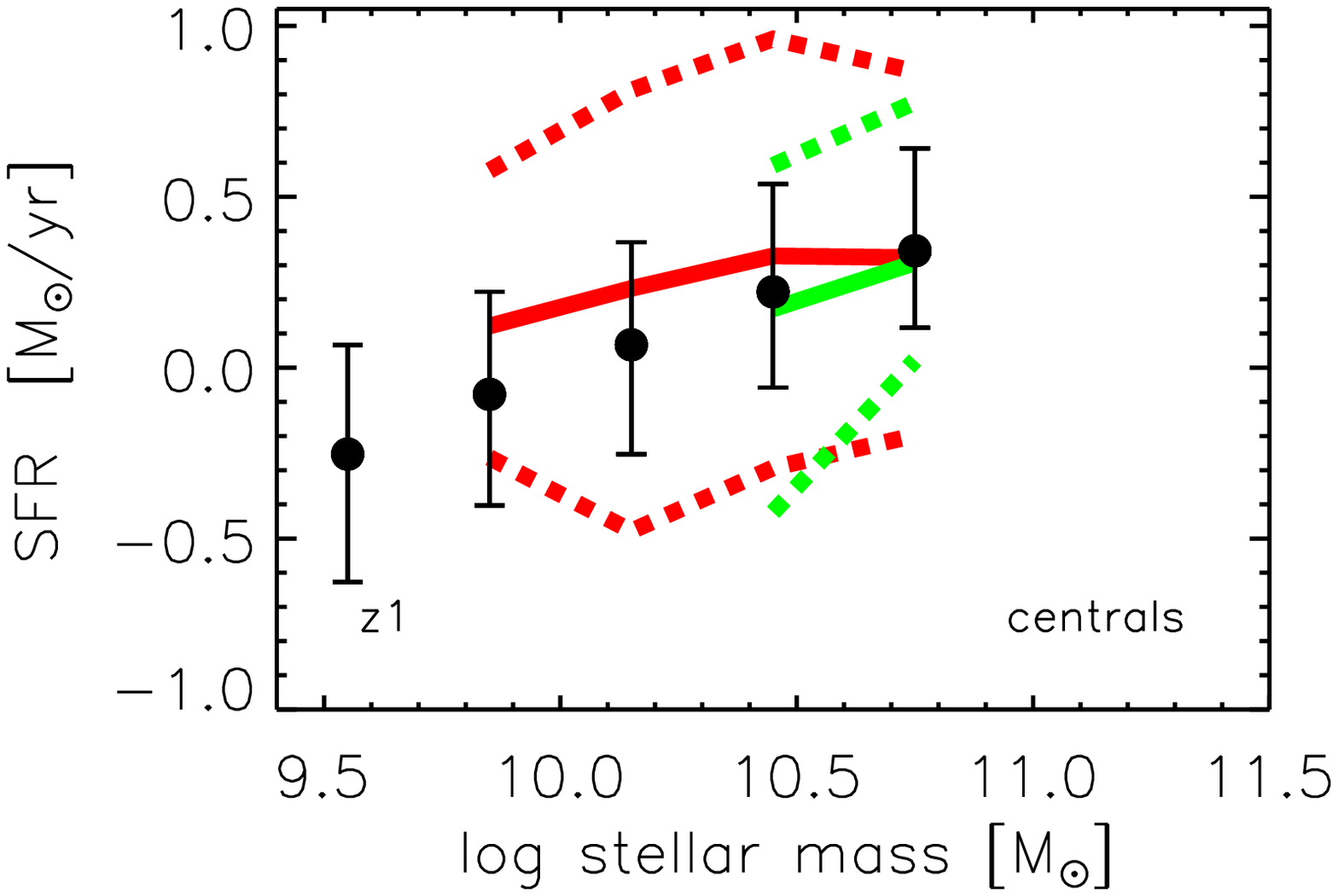}
\includegraphics[height=1.3in,width=1.75in]{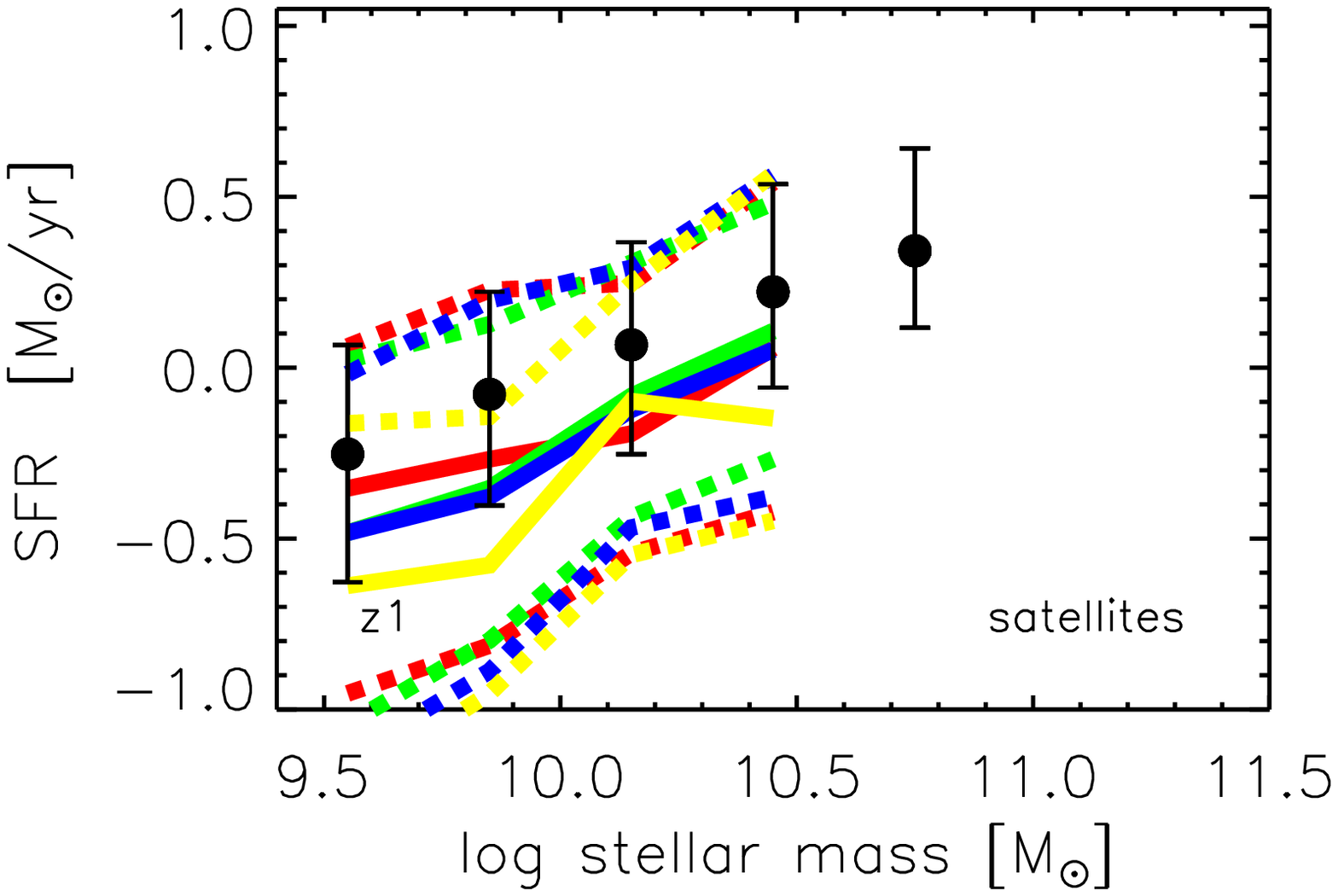}
\includegraphics[height=1.3in,width=1.75in]{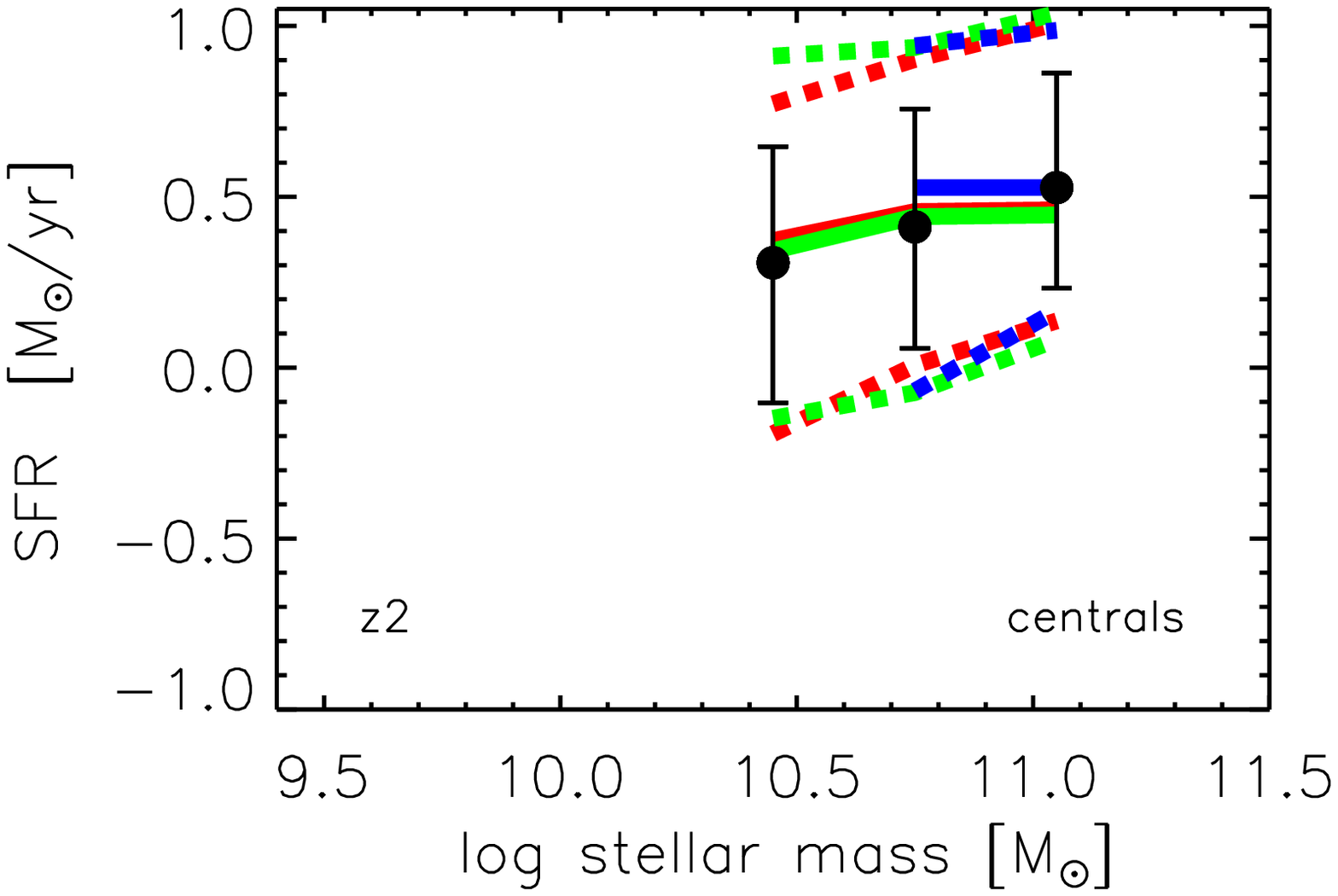}
\includegraphics[height=1.3in,width=1.75in]{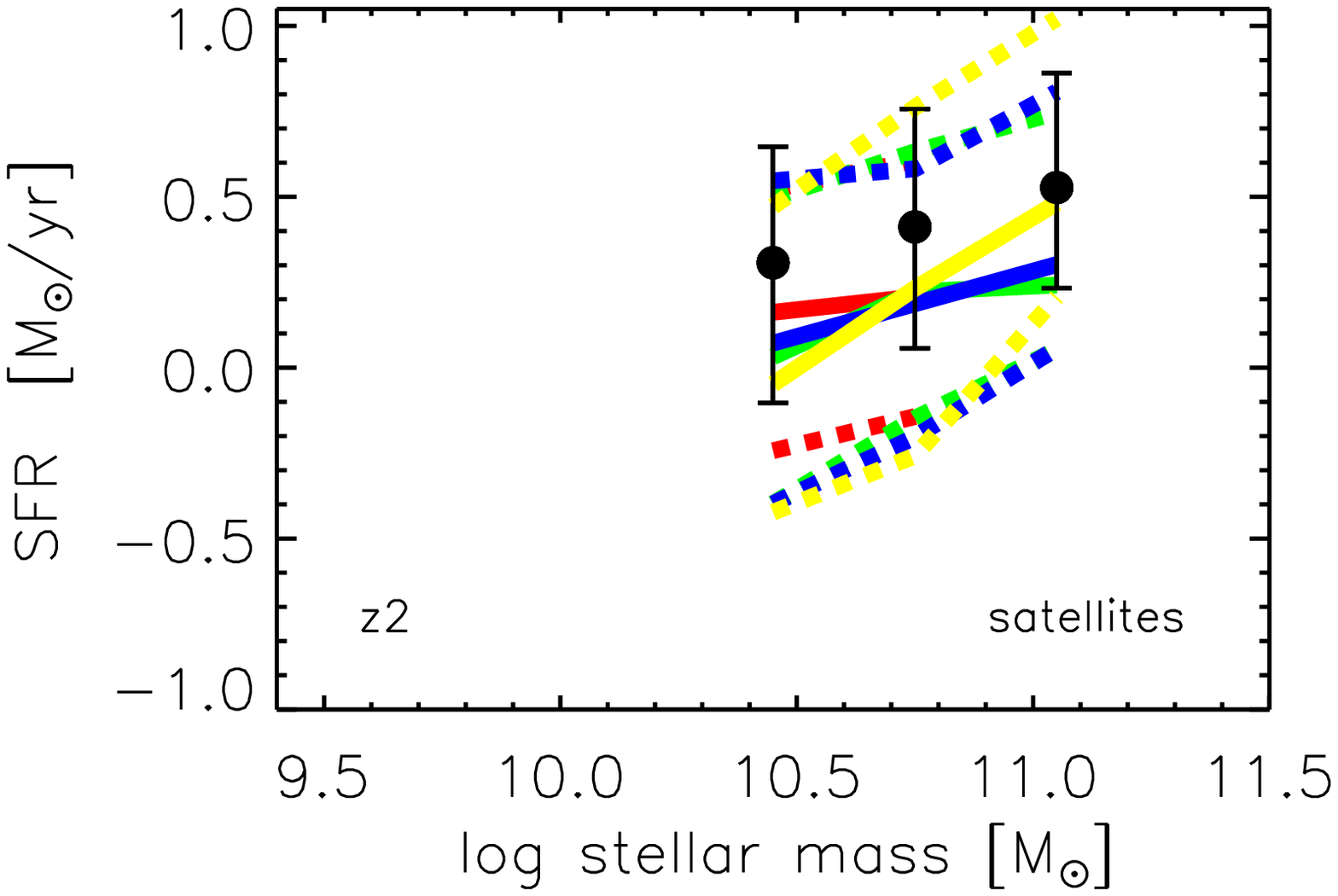}
\includegraphics[height=1.3in,width=1.75in]{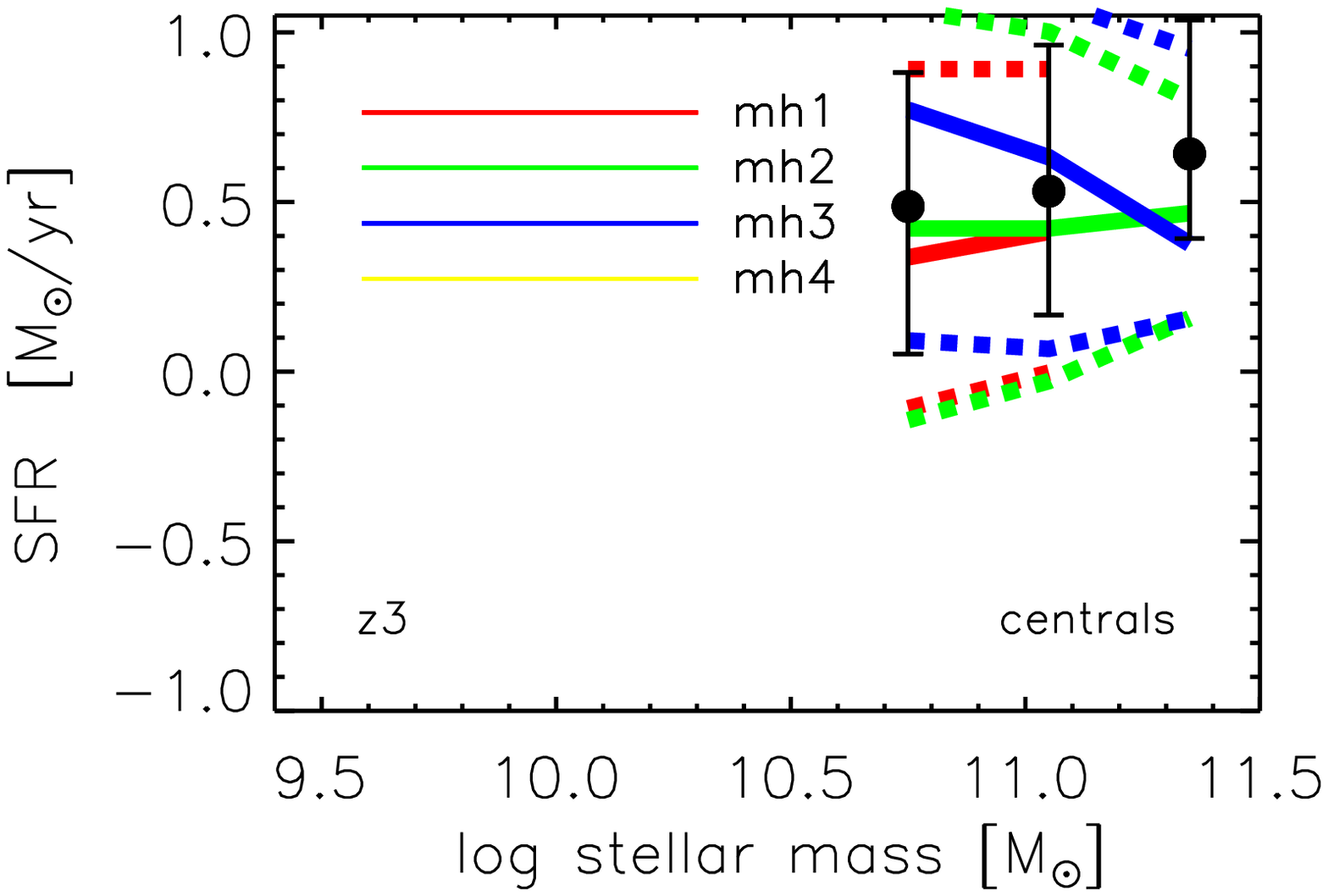}
\includegraphics[height=1.3in,width=1.75in]{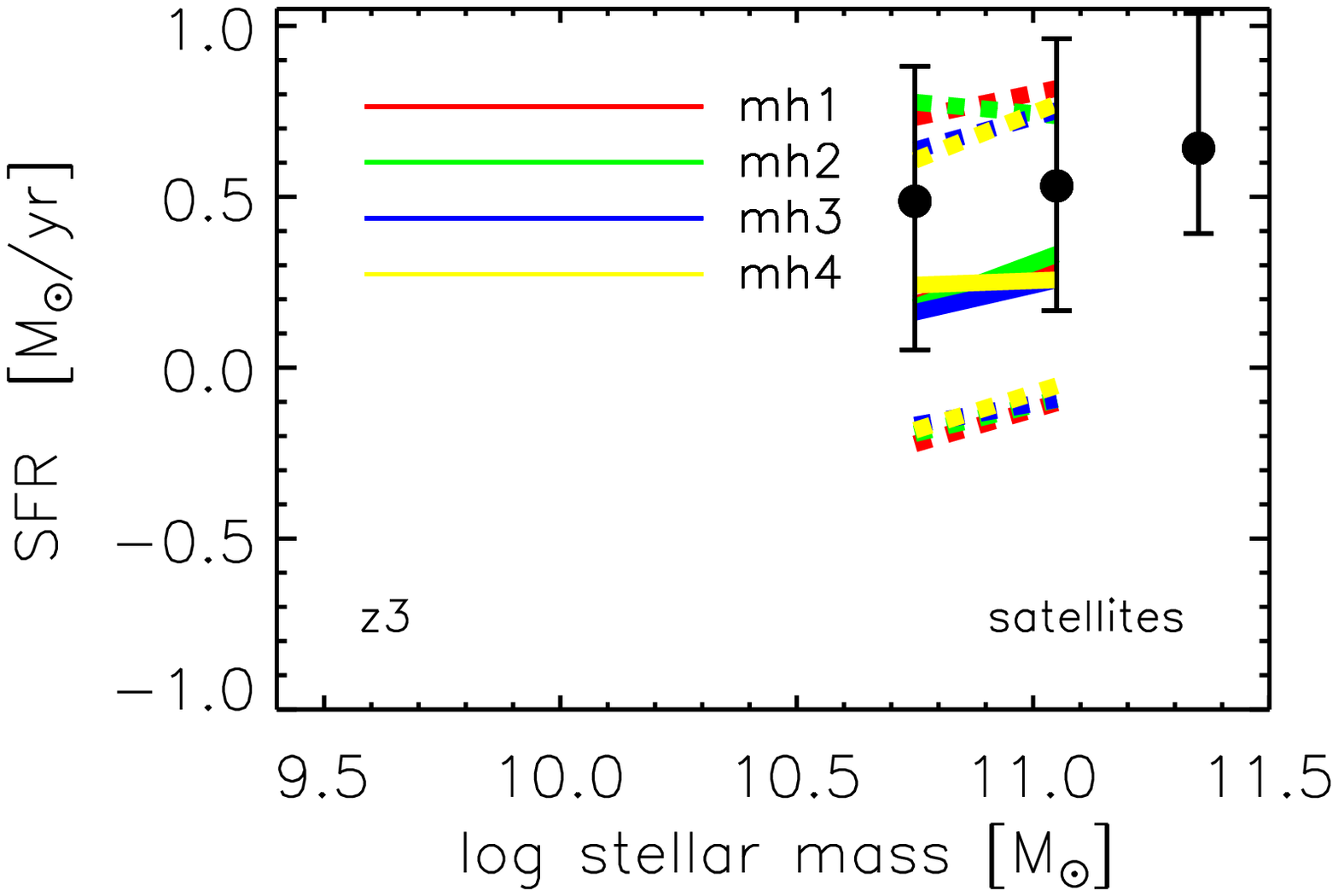}
\caption{The 16th, 50th and 84th percentiles in the SFR of the SFGs as a function of stellar mass (top: $z1$; middle: $z2$; bottom: $z3$). Field galaxies (filled circles) are compared with centrals (left) and satellites (right), divided into four halo mass bins ($mh1=[10^{12}, 10^{12.5}] M_{\odot}$; $mh2=[10^{12.5}, 10^{13}] M_{\odot}$; $mh3=[10^{13}, 10^{13.5}] M_{\odot}$; $mh4=[10^{13.5}, 10^{14}] M_{\odot}$).}
\label{ms}
\end{figure}

We define galaxies with SSFRs greater than the threshold $T$ as star-forming galaxies (SFG). Now, we can study the SFR distribution of the SFG in different halo environments and at different cosmic epochs. Fig. ~\ref{ms} shows the 16th, 50th and 84th percentile in the SFR distribution for either centrals or satellites as a function of stellar mass, compared to field galaxies. For clarity, we use filled dots with "error bars" to represent the 16th, 50th and 84th percentile for field galaxies. Therefore, these "error bars" indicate the width of the SFR distribution (not the uncertainty on the median value). The centrals seem to have very similar median SFR but somewhat wider distributions compared to field galaxies. We do not see strong dependence on halo mass, possibly due to the small sample size and the small dynamic range (in halo mass) or a real absence of environmental dependence for the centrals. For the satellites, it is clear that the median SFR is consistently lower and the width in the SFR distribution is also consistently larger compared to field galaxies. In the lowest redshift bin, there is indication that at fixed stellar mass, the median SFR for the satellites in the most massive halos is somewhat lower (although not consistent across the stellar mass range) compared to satellites in less massive halos. 

\begin{figure}
\centering
\includegraphics[height=1.3in,width=1.75in]{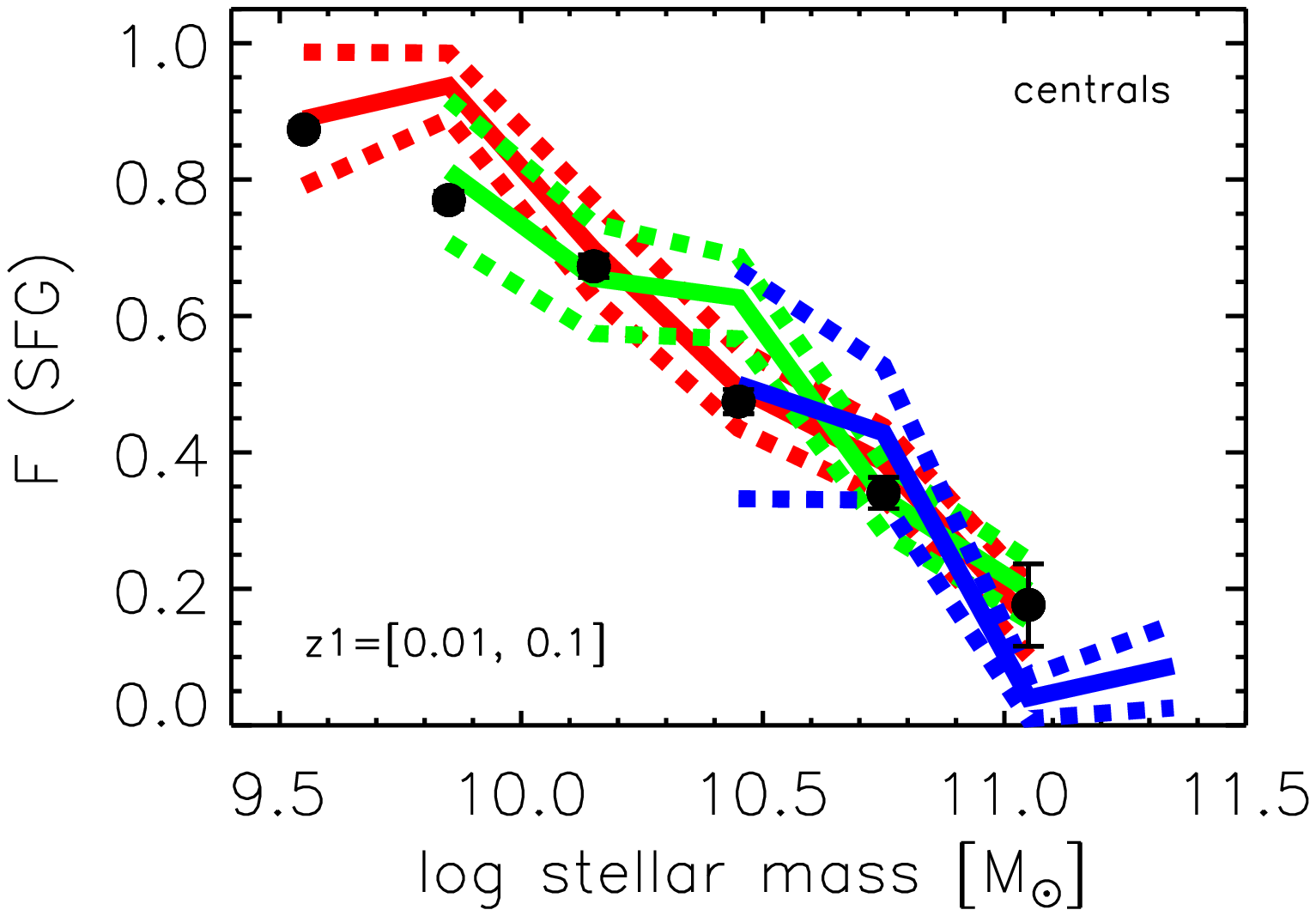}
\includegraphics[height=1.3in,width=1.75in]{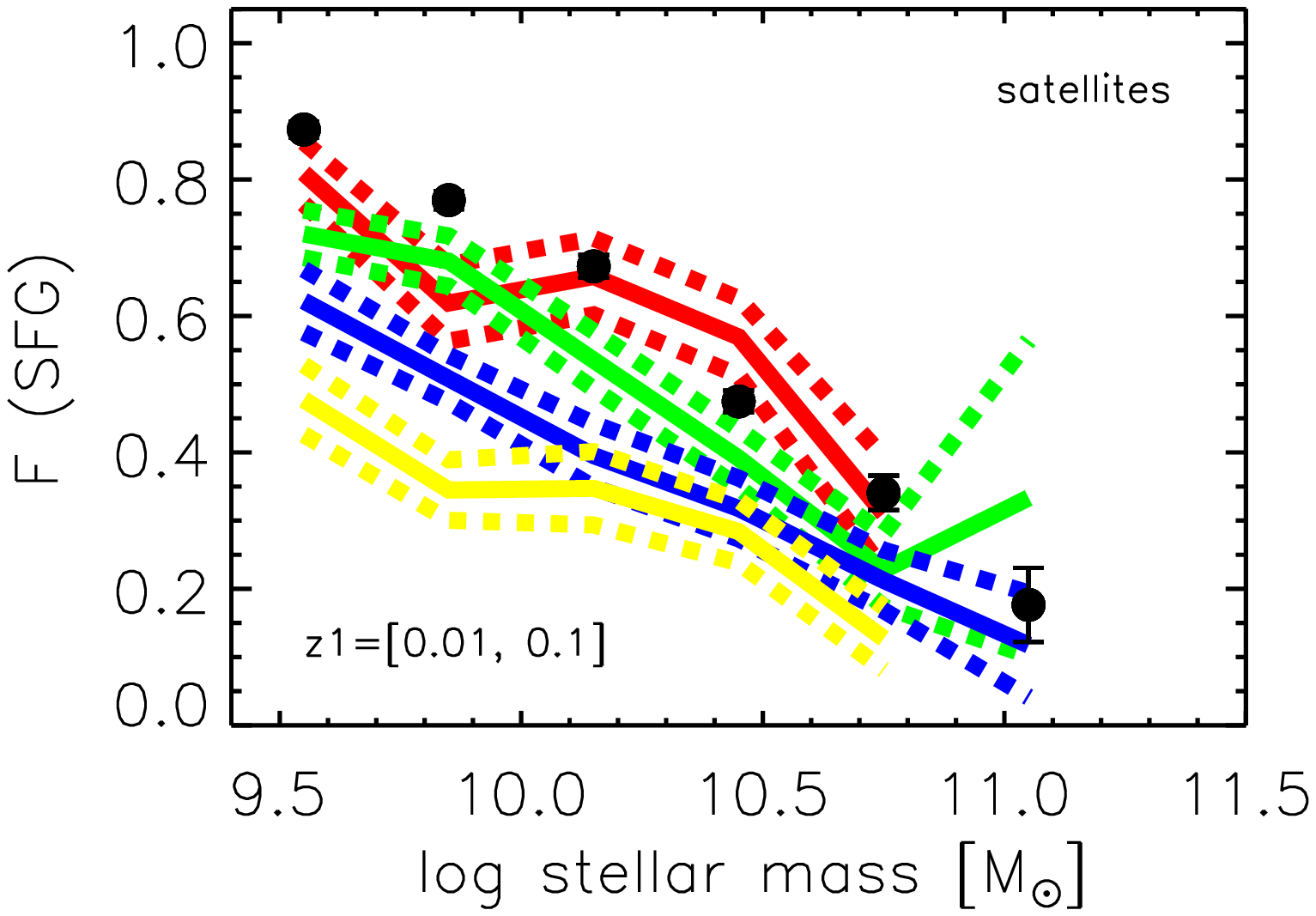}
\includegraphics[height=1.3in,width=1.75in]{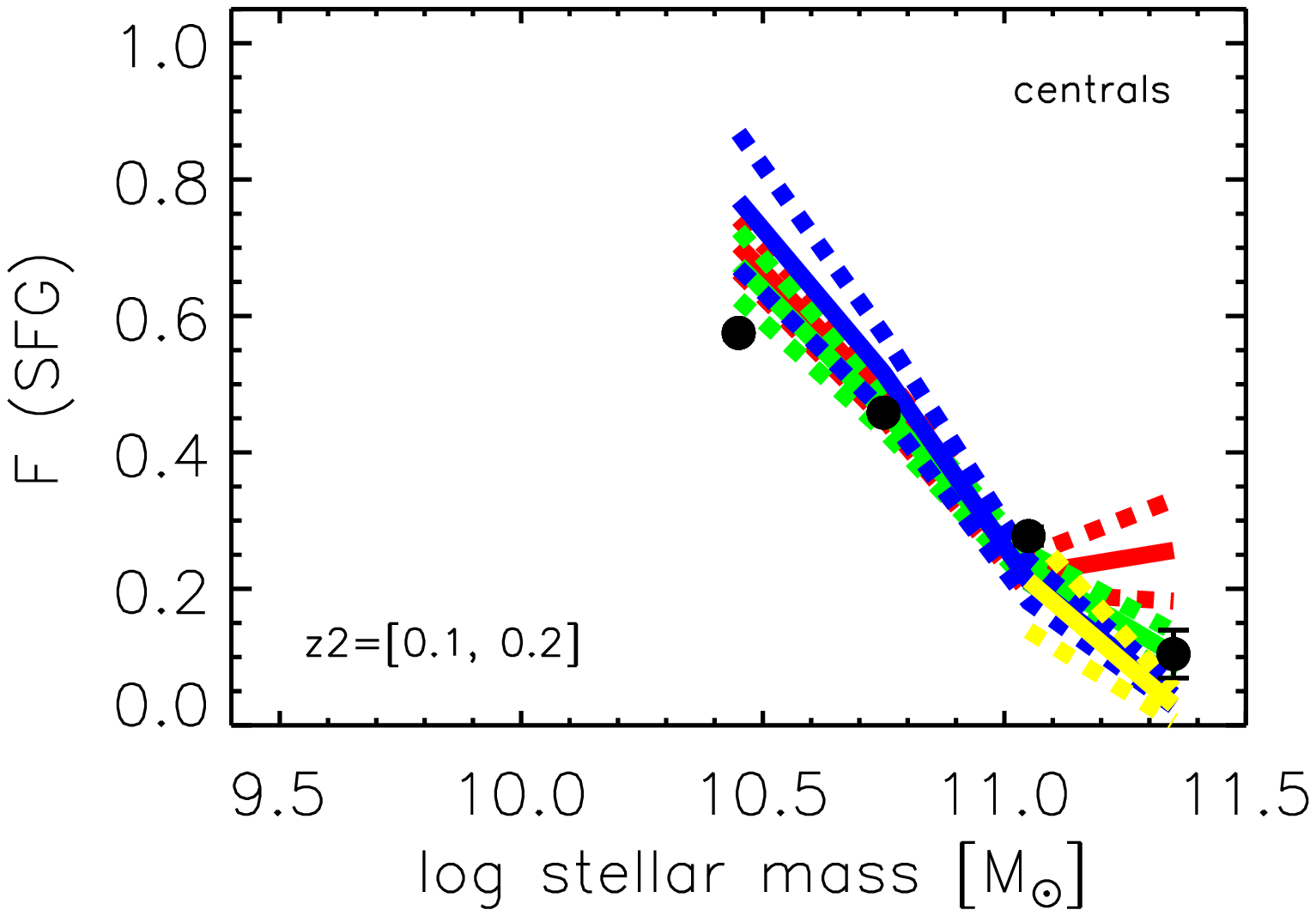}
\includegraphics[height=1.3in,width=1.75in]{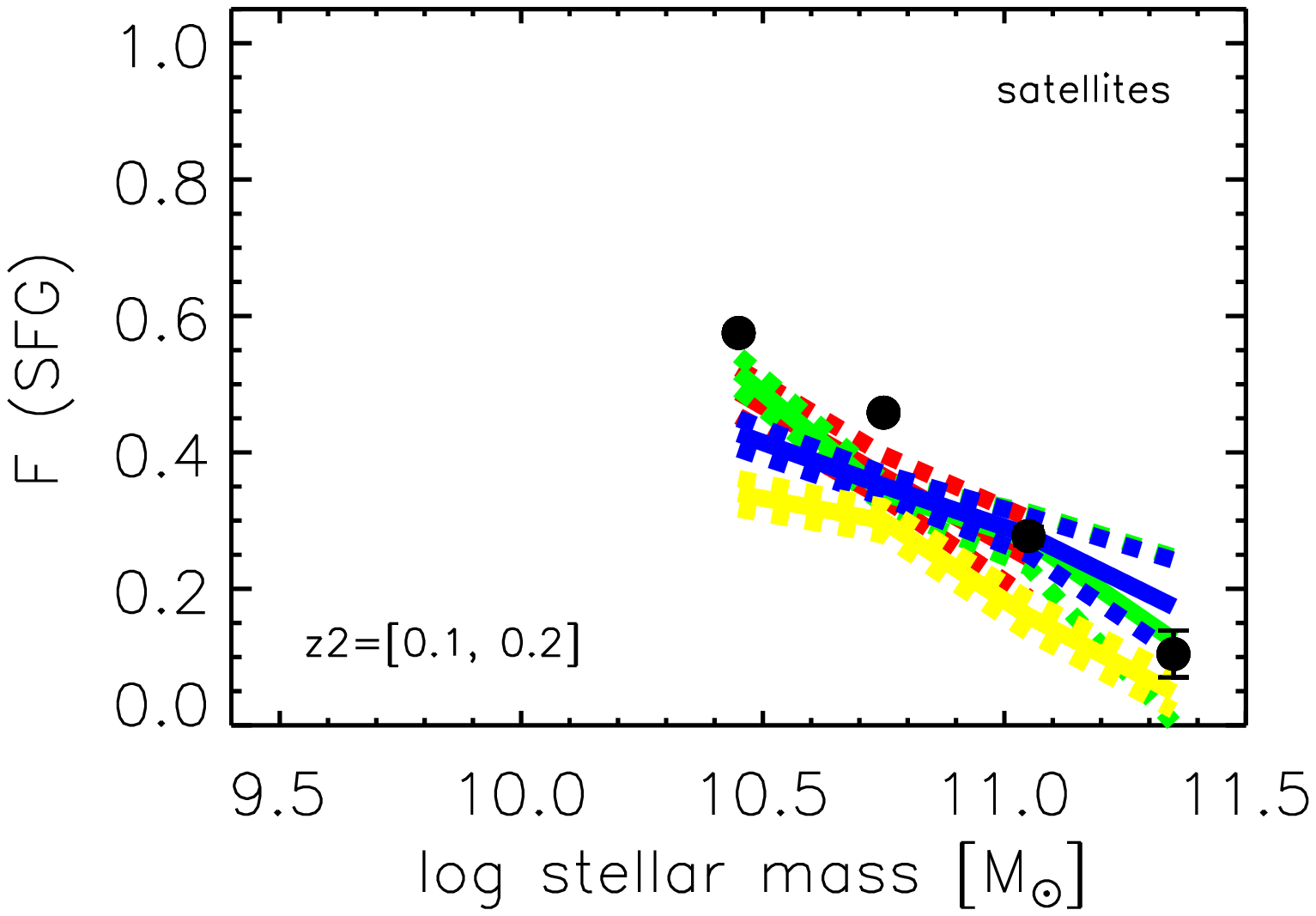}
\includegraphics[height=1.3in,width=1.75in]{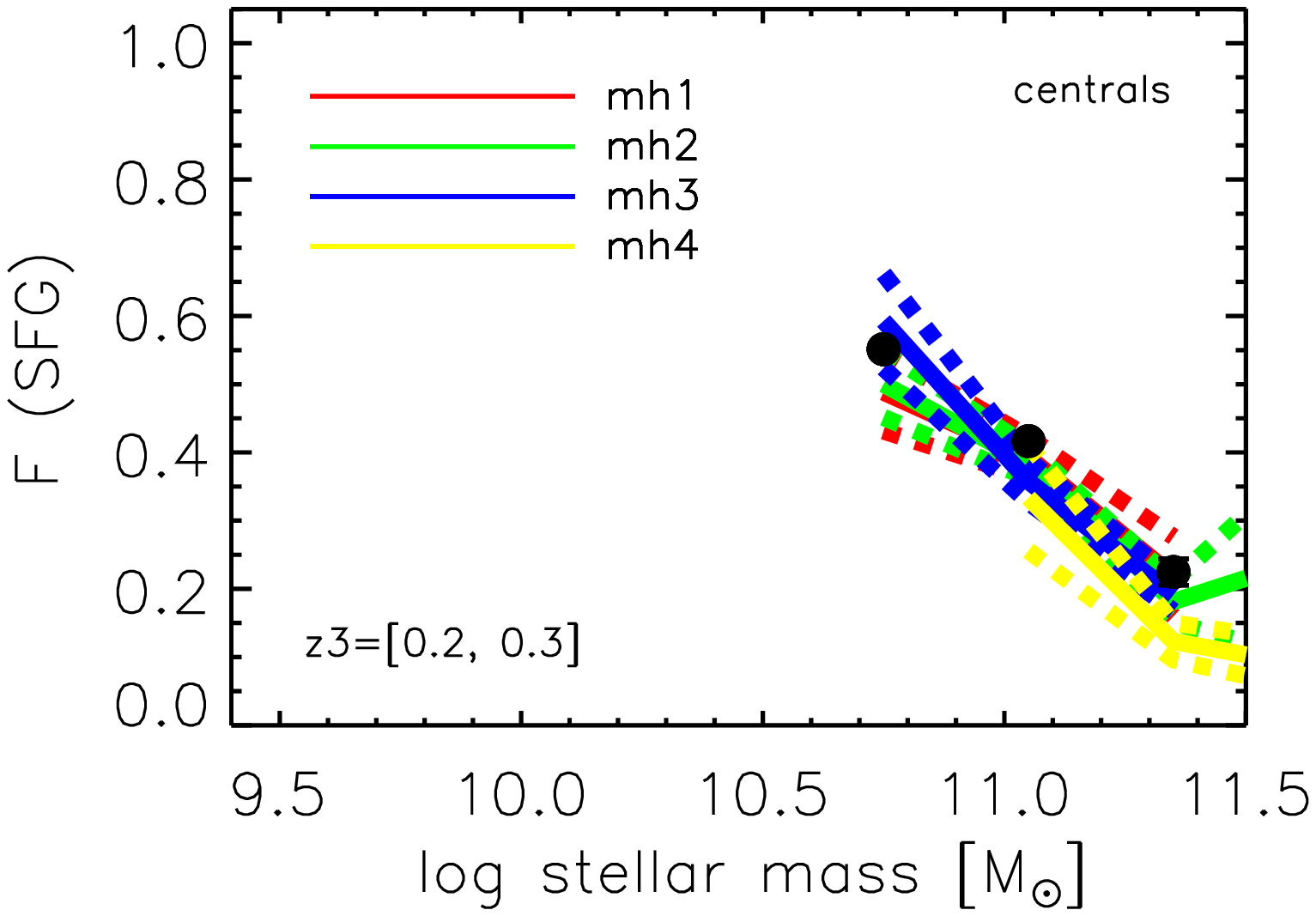}
\includegraphics[height=1.3in,width=1.75in]{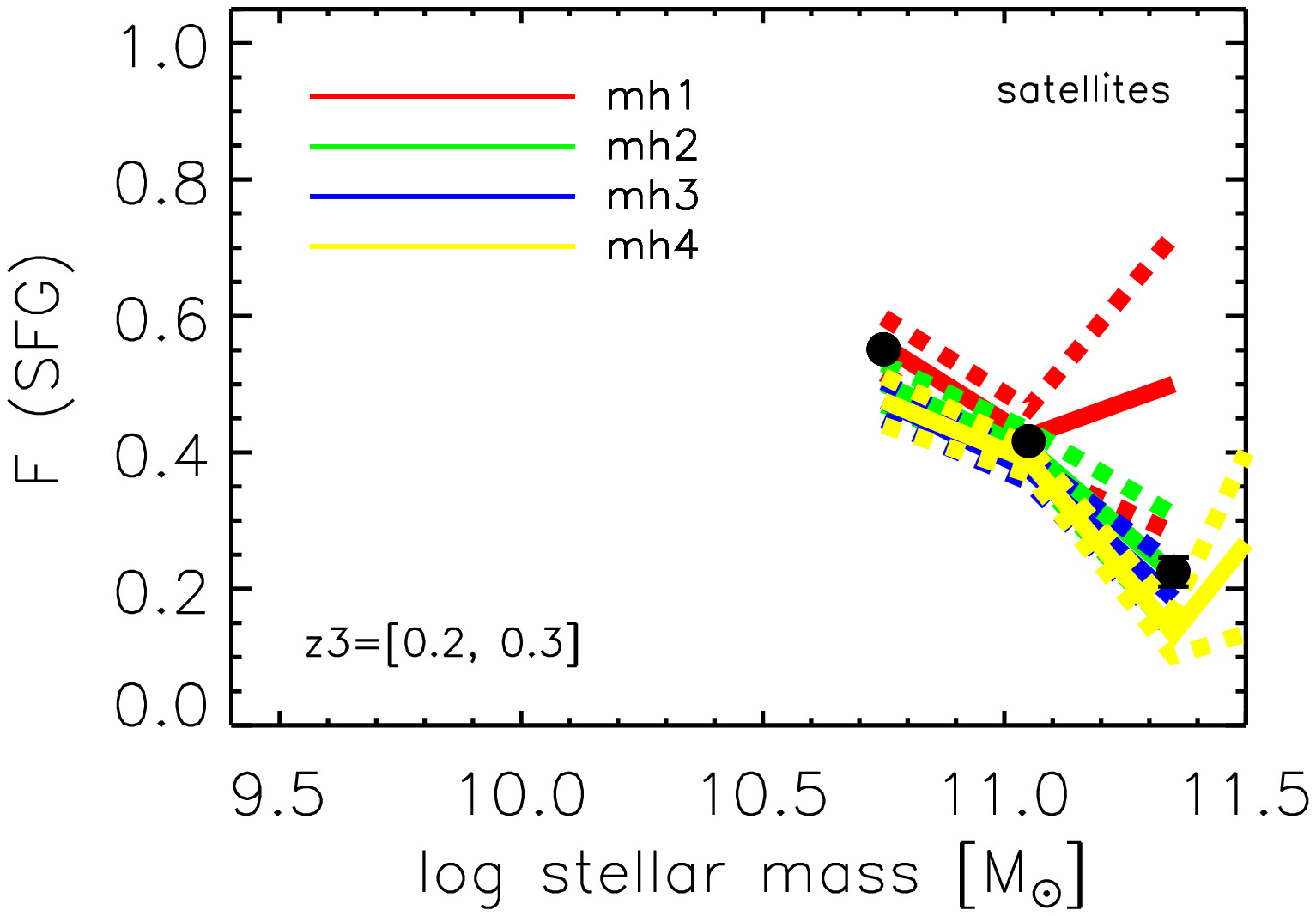}
\caption{${\rm F(SFG)}$ as a function of stellar mass (top: $z1$; middle: $z2$; bottom: $z3$). Field galaxies (filled circles) are compared with centrals (left) and satellites (right) in four halo mass bins.}
\label{fsf}
\end{figure}

Fig.~\ref{fsf} shows the fraction of SFG, ${\rm F(SFG)}$, (the ratio of the number of SFGs over the total number of galaxies) as a function of stellar mass for either centrals or satellites compared to the field galaxies. For centrals, ${\rm F(SFG)}$ in different halo mass bins is always similar to ${\rm F(SFG)}$ in the field which suggests that the environmental processes (represented as the fraction of SFG) quenching star formation do not affect centrals. For satellites, ${\rm F(SFG)}$  as a function of stellar mass exhibits strong variation in halos of different mass ranges with ${\rm F(SFG)}$ increasing as halo mass decreases. However, in the highest redshift bin, we are unable to detect variation between satellites and field galaxies and any dependence on halo mass due to small sample size and small dynamic range. Another possibility is that the dependence of ${\rm F(SFG)}$ on halo mass weakens with increasing redshift and at $z>0.2$ it is barely noticeable. In halos with masses in the range $[10^{12}, 10^{12.5}]$, ${\rm F(SFG)}$ for the satellites is similar to field galaxies. The difference in ${\rm F(SFG)}$ for satellites over the halo mass range between $10^{12}M_{\odot}$ and $10^{14}M_{\odot}$ increases significantly over the last 2 Gyr or so since $z\sim0.2$.

 \subsection{Stellar mass dependence}
 
Our definition of SFG depends on galaxy bimodality, i.e., how galaxies separate into two classes in the SSFR distribution. It is possible that the details of the bimodal distribution (e.g., the precise value of $T$) also depend on stellar mass, in addition to redshift and galaxy type (field galaxies, centrals or satellites). To address potential dependence on stellar mass, we perform a two Gaussian fit to the $\log$ SSFR distribution in bins of stellar mass and redshift for each galaxy type (shown in Appendix A). 
 
Fig.~\ref{BIMO} shows the parameters of the two Gaussian fits to the $\log$ SSFR distribution as a function of stellar mass. The filled black dots show the location parameters of the star-forming Gaussian for field galaxies. The filled black triangles show the location parameters of the quiescent Gaussian for field galaxies. The "error bars" on the plotting symbols represent the scale parameters (i.e. the width of the Gaussian) and not the uncertainty on the mean (i.e. the location parameter). The lines represent group galaxies. For the centrals or satellites, the middle line corresponds to the location parameter of the Gaussian and the two outer lines correspond to the scale parameter. We can see that for the star-forming Gaussian, satellites have consistently lower mean $\log$ SSFR values (by $\sim0.2$ dex) and a wider spread. In comparison, centrals are much more similar to field galaxies. These trends are in agreement with what is presented in Fig. 3. For the quiescent Gaussian, field galaxies and group galaxies (centrals or satellites) are quite similar to each other (in terms of mean and scatter). The similarity in the quiescent Gaussian in the field and group environment as well as in the central and satellite population may suggest that once a galaxy is quenched, it becomes a quiescent galaxy of fairly homogeneous properties insensitive to its environment. Our findings of the environmental dependence of the positions of the MS for centrals and satellites are similar to conclusions in Grootes et al. (2017) for morphologically selected spirals.

In Fig.~\ref{fsf_mstar}, we plot the fraction of SFG ${\rm F(SFG)}$ as a function of stellar mass and examine how it depends on halo mass, redshift and galaxy type. Note that the difference between Fig.~\ref{fsf} and Fig.~\ref{fsf_mstar} is that the threshold used to select SFG is independent of stellar mass in Fig.~\ref{fsf}. Despite this difference, we can draw similar conclusions. We see little variation in ${\rm F(SFG)}$ between field galaxies and centrals and little halo mass dependence, although we are limited by the small sample size and small range in stellar mass. In contrast, ${\rm F(SFG)}$ for satellites depends strongly on halo mass in the lowest redshift bin with decreasing ${\rm F(SFG)}$ at a given stellar mass as halo mass increases. From halos in the mass range $[10^{12}, 10^{12.5}] M_{\odot}$ to halos in the mass range $[10^{13.5}, 10^{14}] M_{\odot}$, ${\rm F(SFG)}$ decreases by roughly a factor of 2. The halo mass dependence is still present in $z2$ but is significantly weaker. It is reasonable to expect that if environmental processes quenching star formation activities in satellites take place on timescale comparable to the time span probed by our redshift bins, then we would see a noticeable increase in ${\rm F(SFG)}$. In the highest redshift bin, our data is insufficient to detect any potential halo mass dependence in the satellite population. It could also indicate that environmental effects on ${\rm F(SFG)}$ in the satellites disappear above $z\sim0.2$.  Our results on the environmental dependence of ${\rm F(SFG)}$ for centrals and satellites are qualitatively similar to the findings in Kova{\v c} et al. (2014).
 
\begin{figure}
\centering
\includegraphics[height=1.3in,width=1.75in]{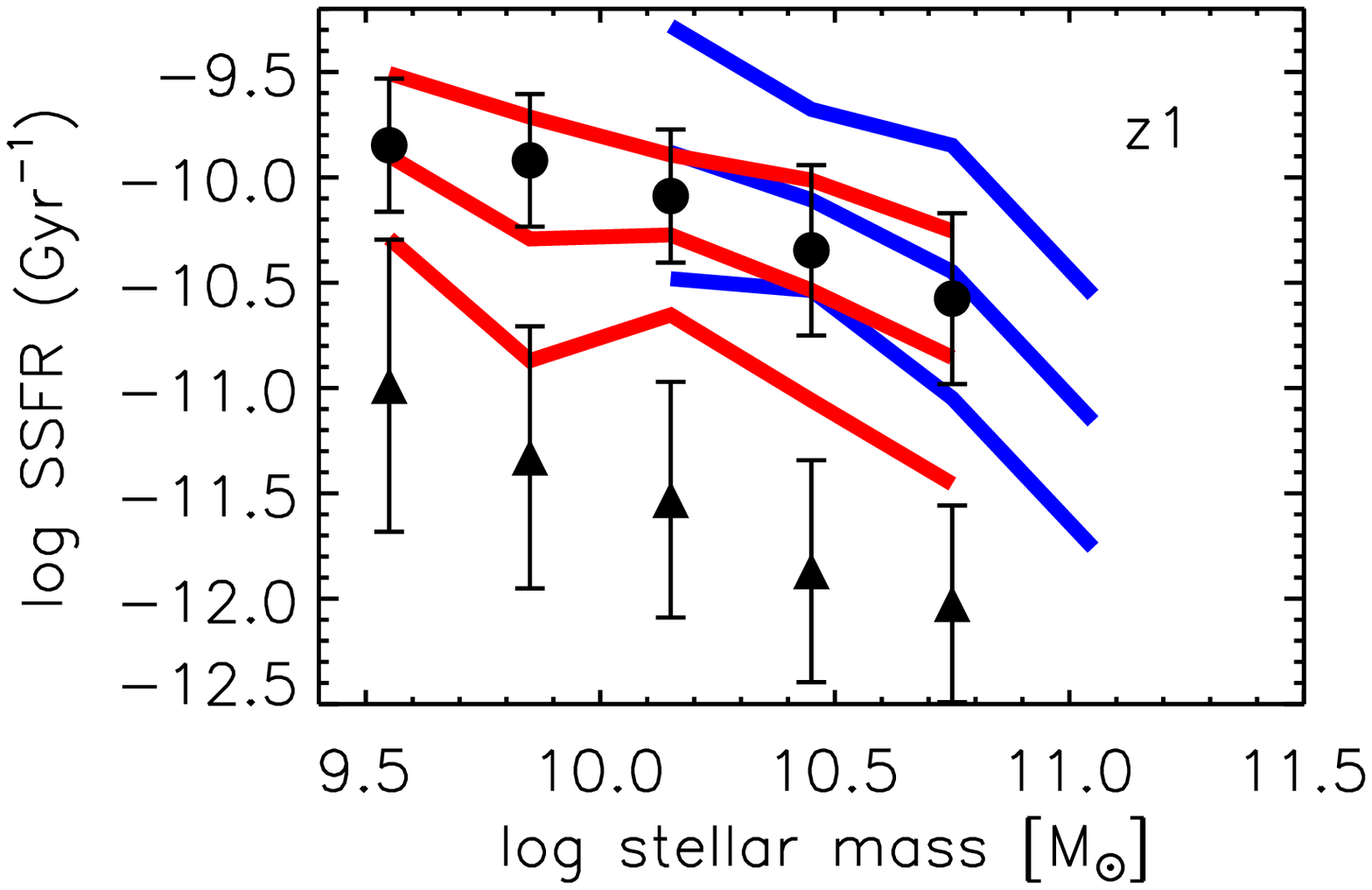}
\includegraphics[height=1.3in,width=1.75in]{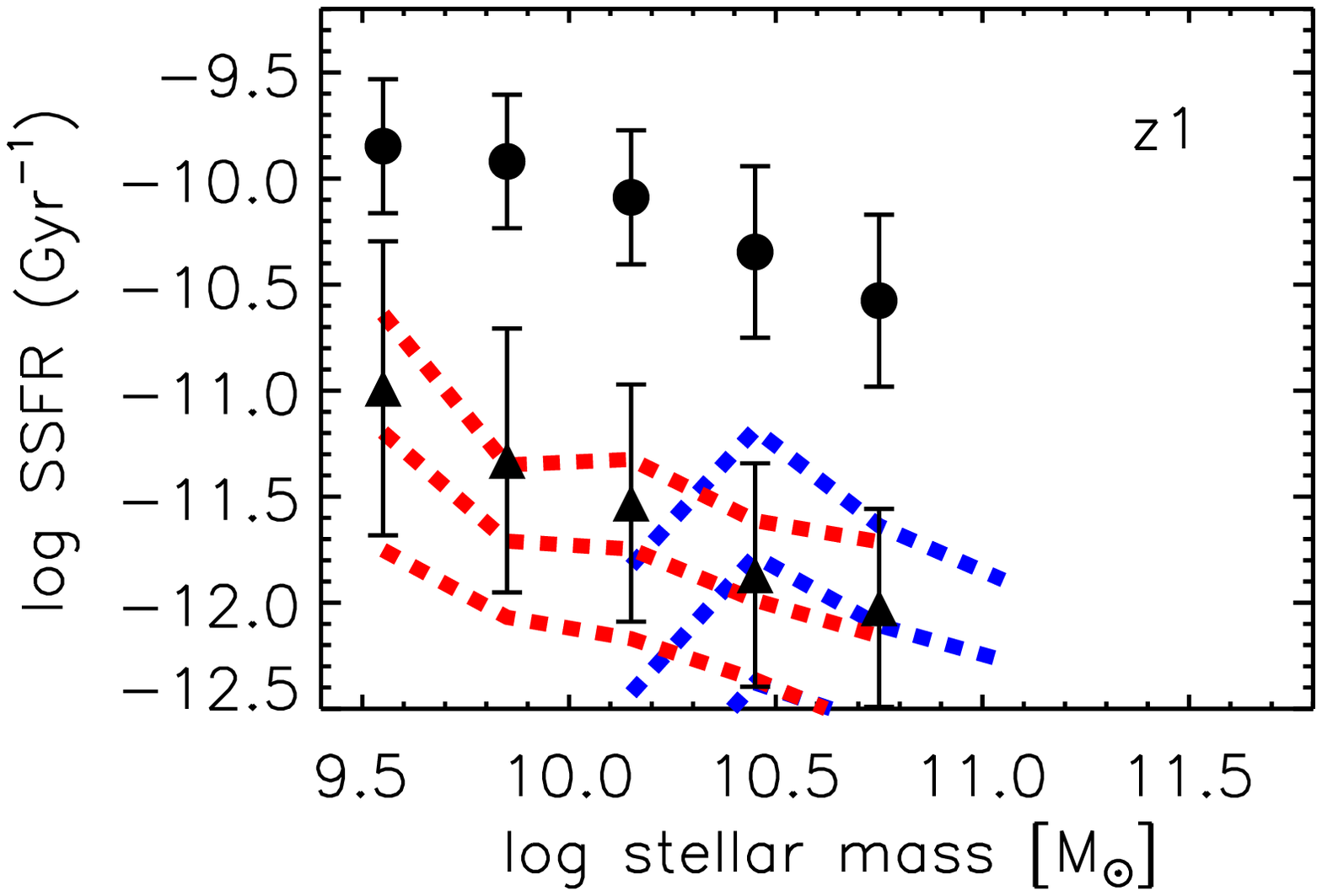}
\includegraphics[height=1.3in,width=1.75in]{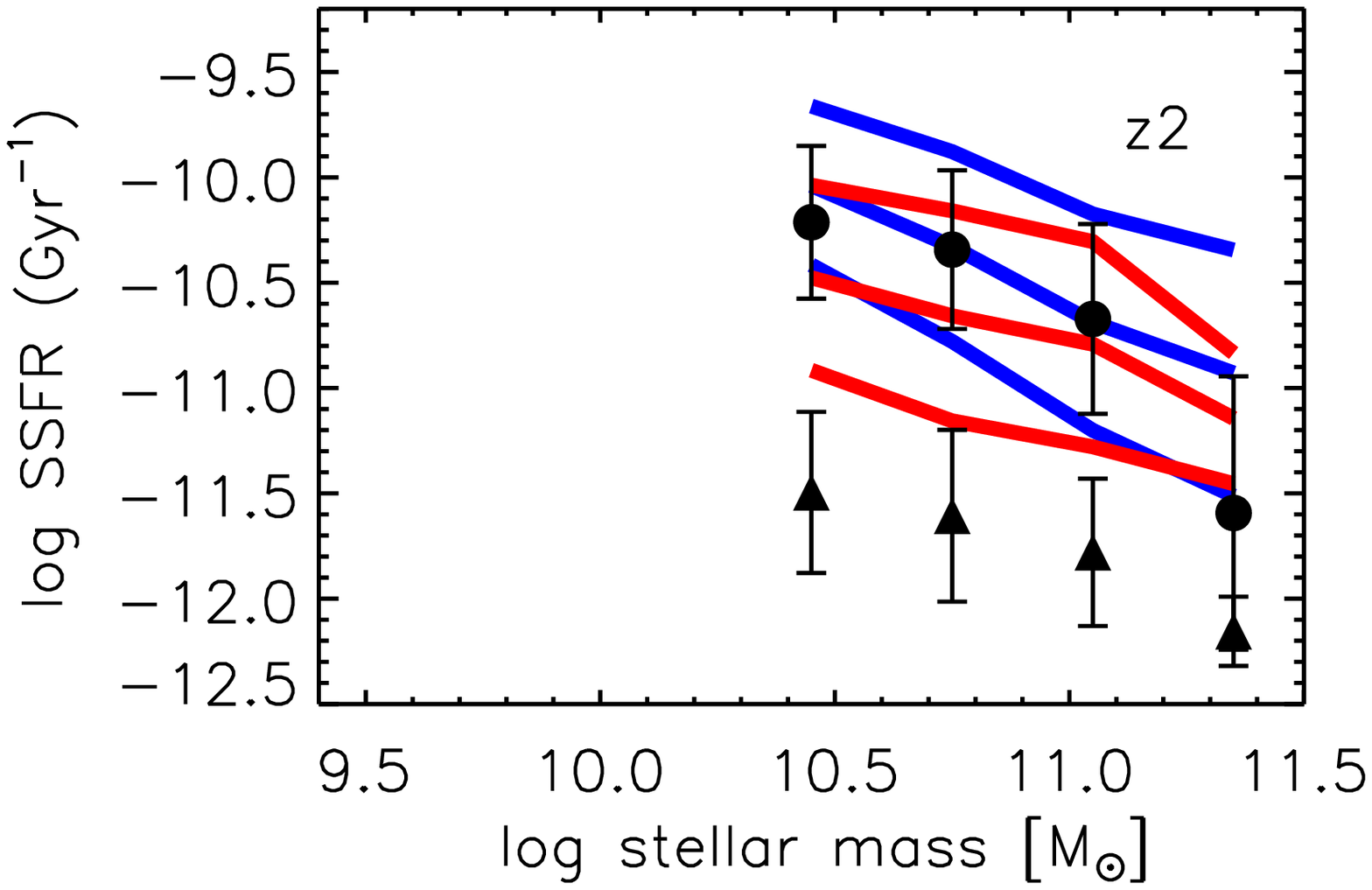}
\includegraphics[height=1.3in,width=1.75in]{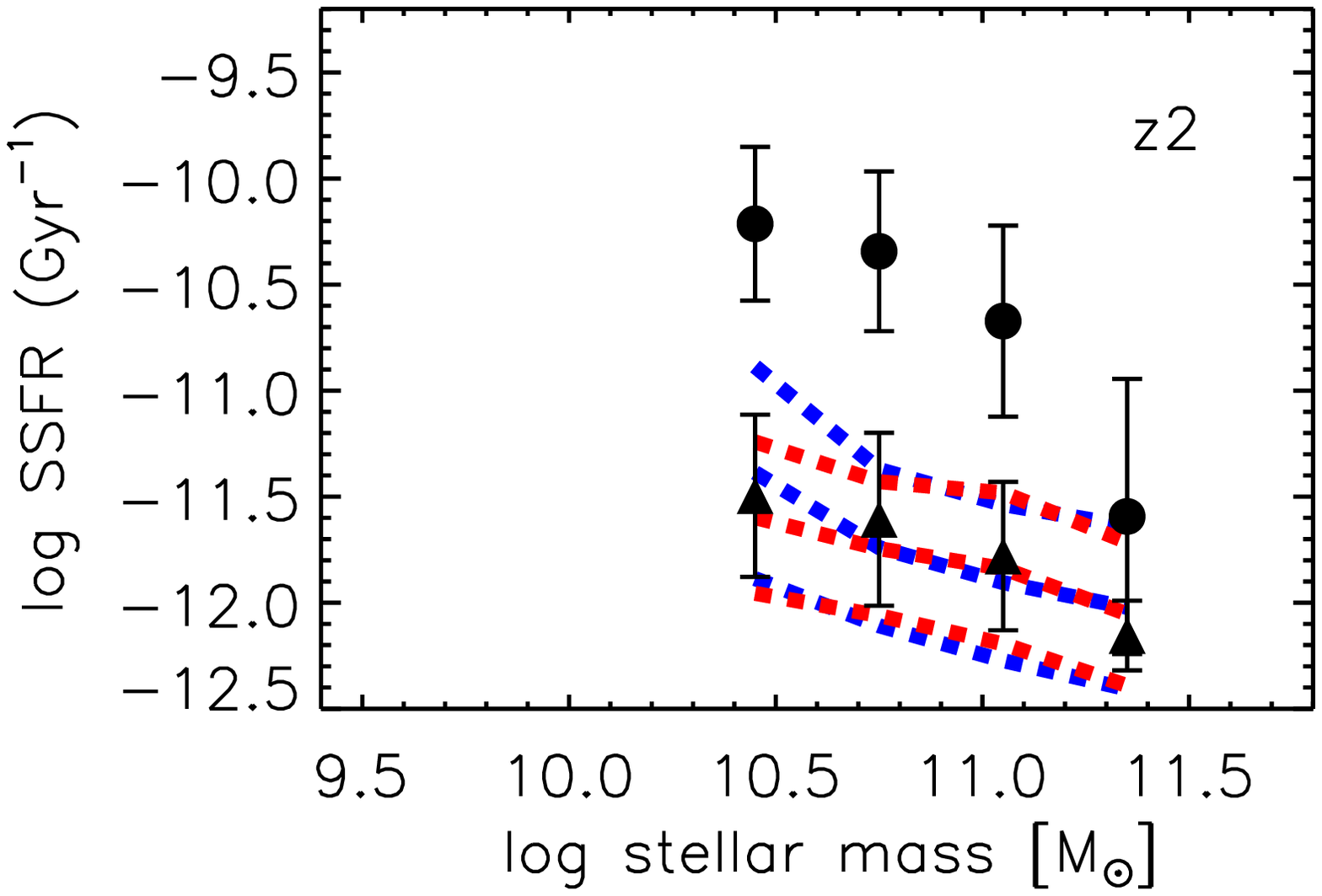}
\includegraphics[height=1.3in,width=1.75in]{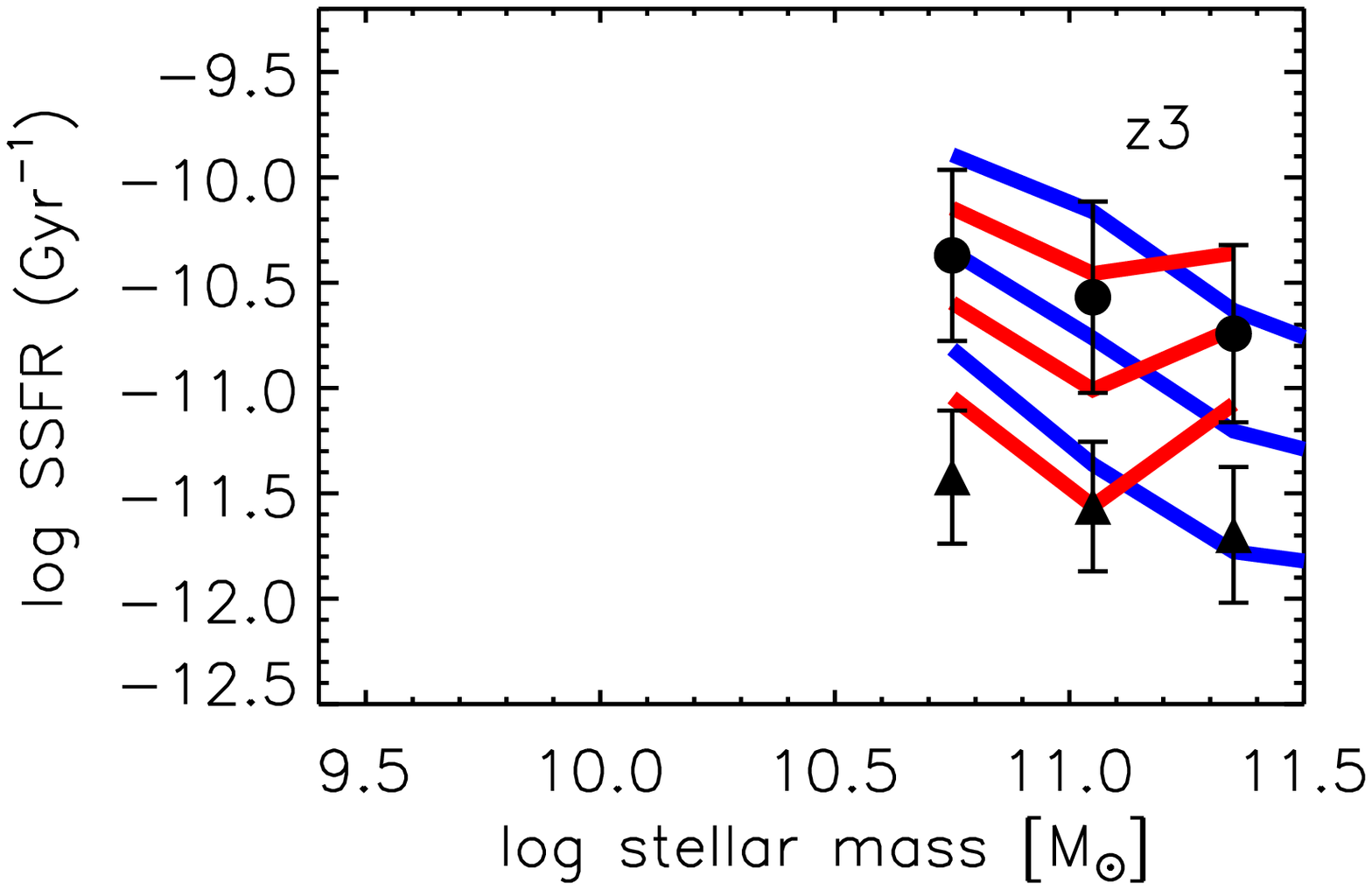}
\includegraphics[height=1.3in,width=1.75in]{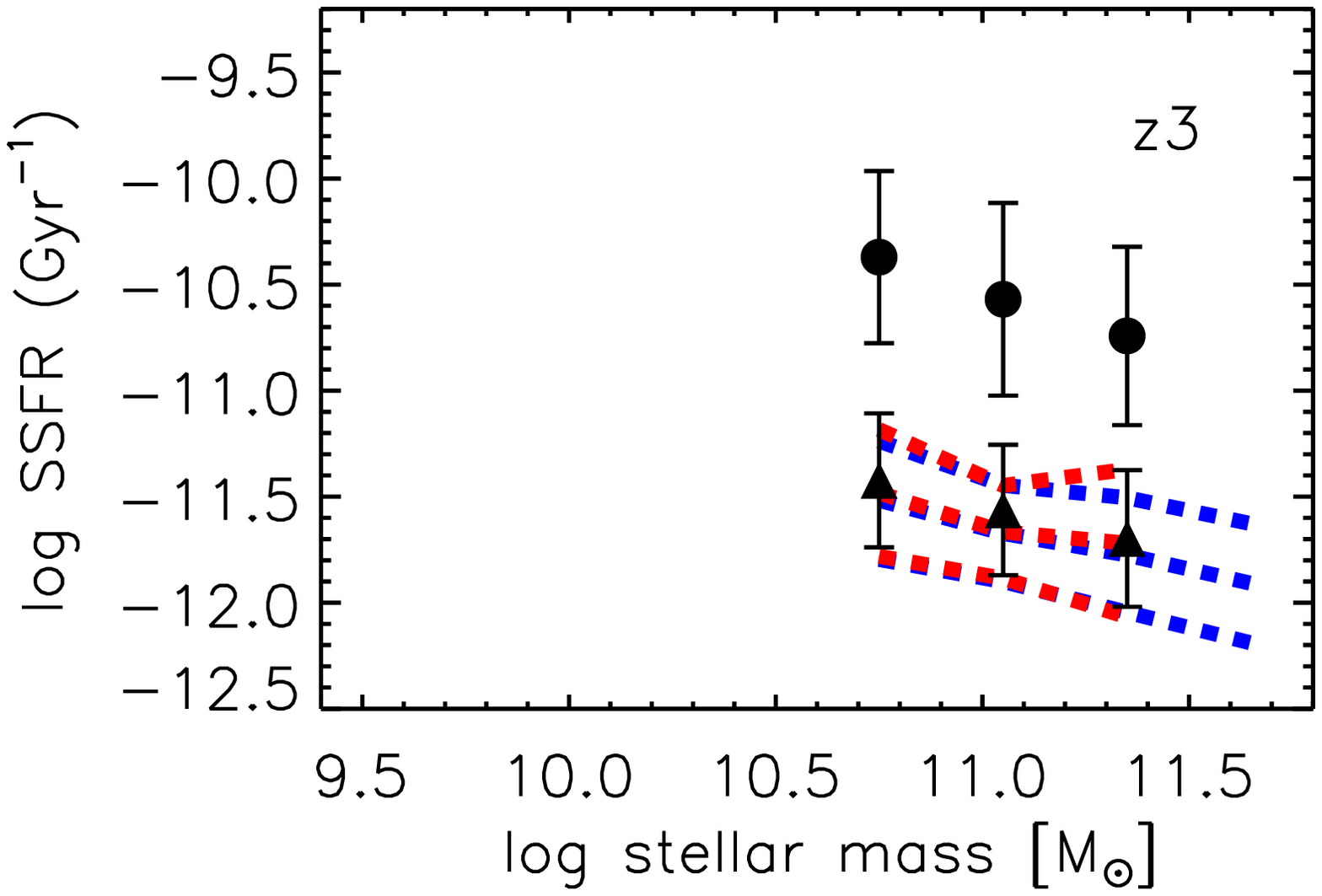}
\caption{Parameters of the Gaussian fits to the log SSFR distribution as a function of stellar mass. The circles show the location parameters of the star-forming Gaussian and the triangles show the location parameters of the quiescent Gaussian for field galaxies. The error bars represent the scale parameters. Left: the location and scale parameters of the star-forming Gaussian for the centrals (blue solid lines) and the satellites (red solid lines) compared to field galaxies. Right: the location and scale parameters of the quiescent Gaussian for the centrals (blue dashed lines) and the satellites (red dashed lines) compared to field galaxies.}
\label{BIMO}
\end{figure}

\begin{figure}
\centering
\includegraphics[height=1.3in,width=1.75in]{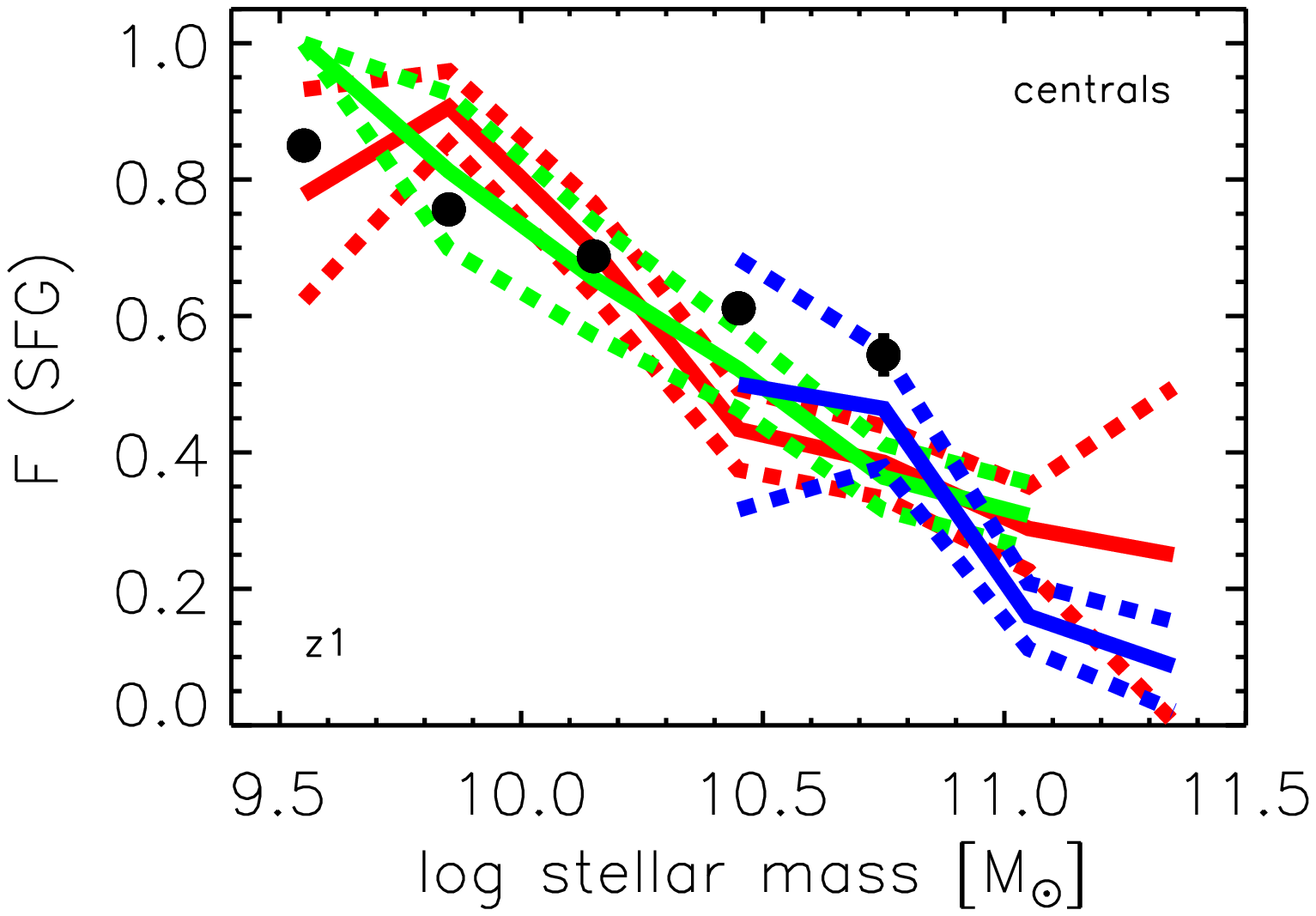}
\includegraphics[height=1.3in,width=1.75in]{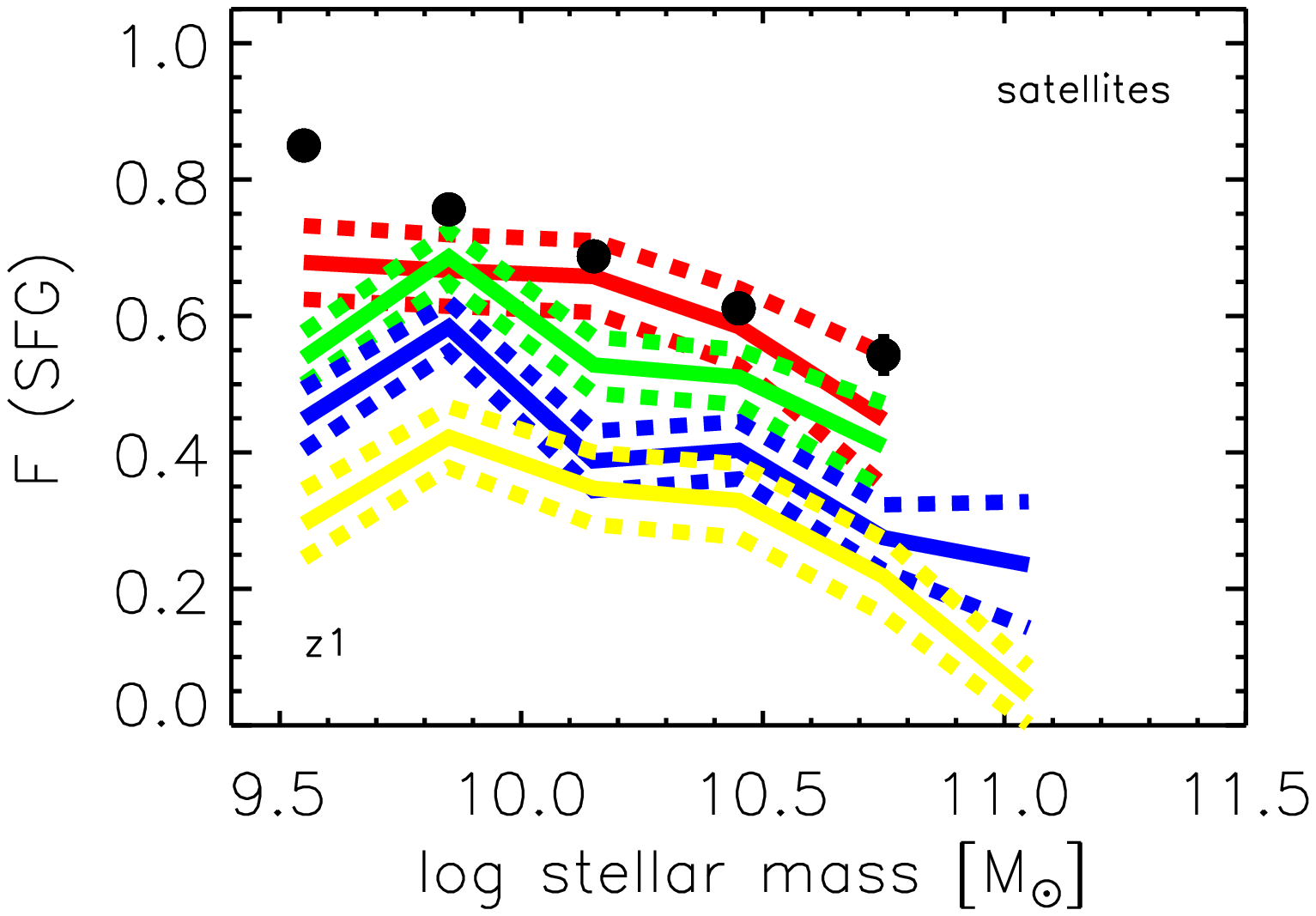}
\includegraphics[height=1.3in,width=1.75in]{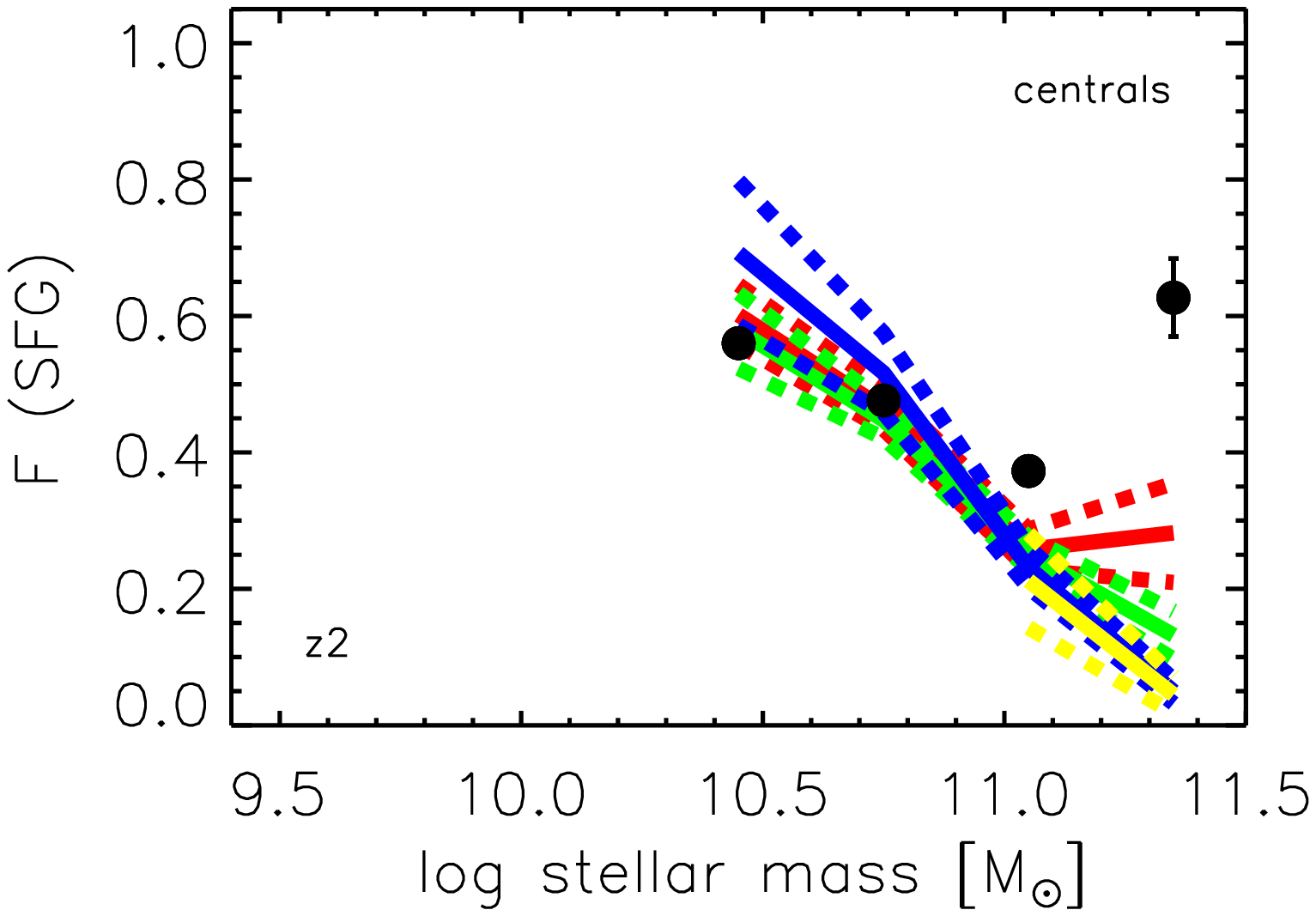}
\includegraphics[height=1.3in,width=1.75in]{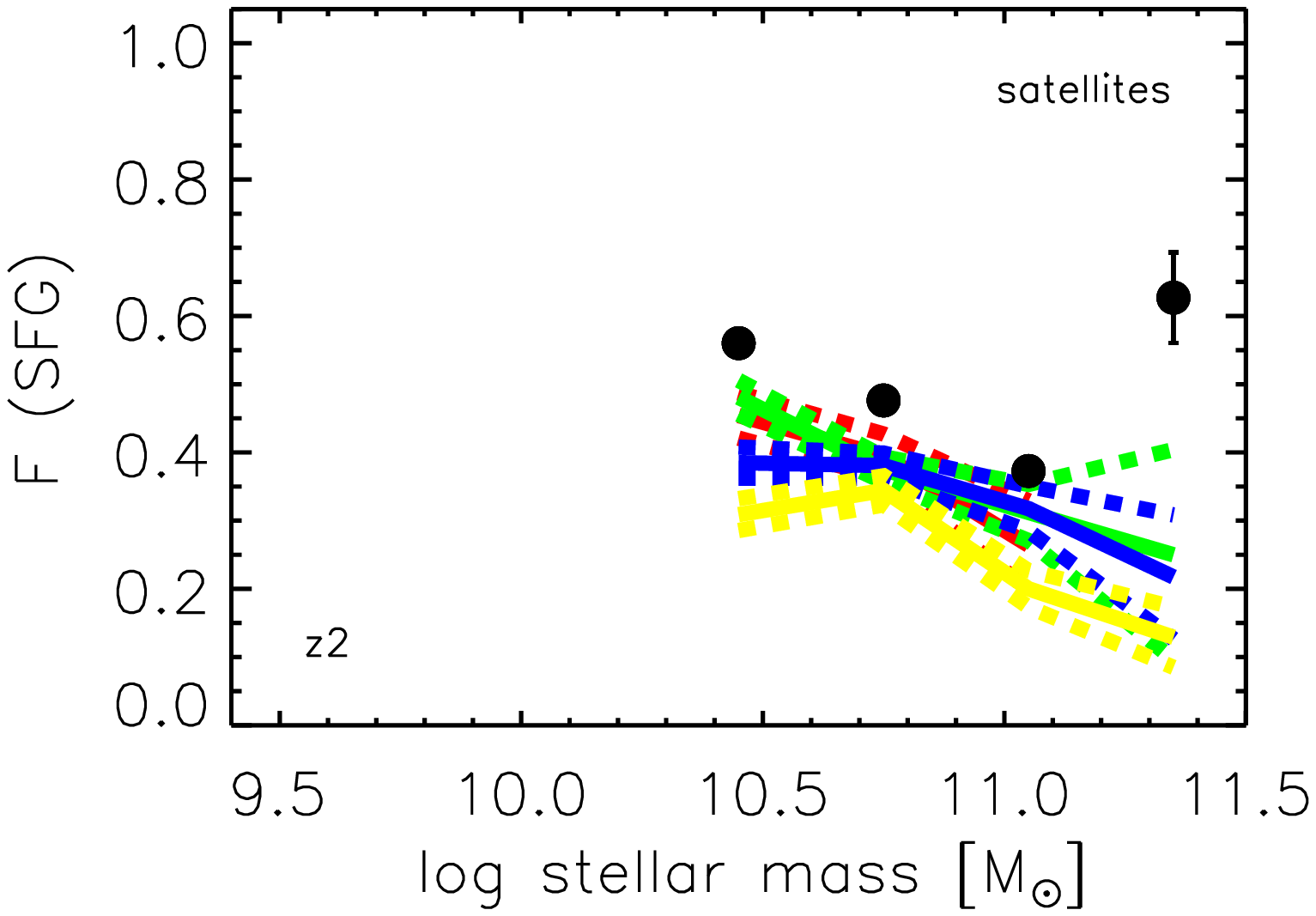}
\includegraphics[height=1.3in,width=1.75in]{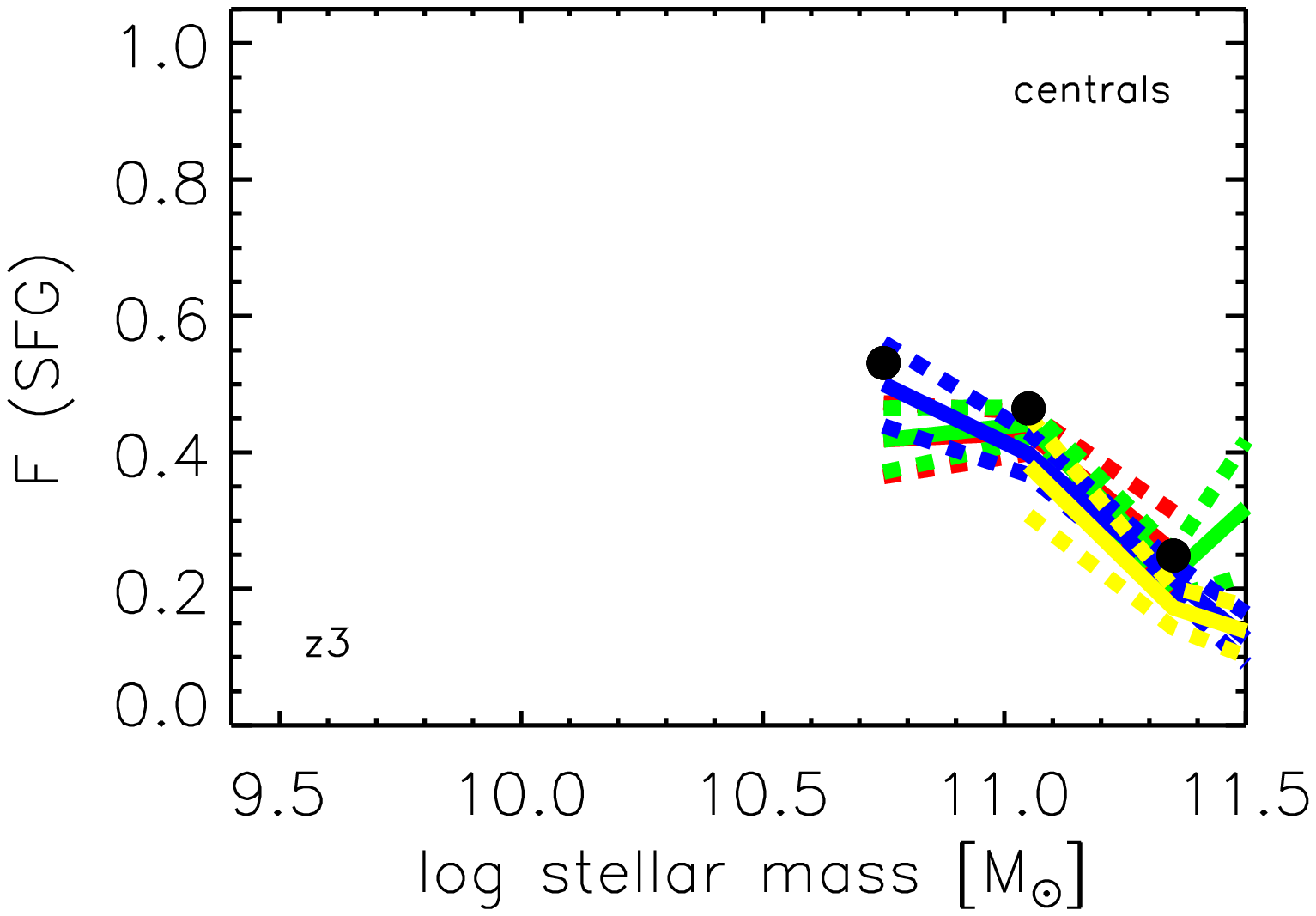}
\includegraphics[height=1.3in,width=1.75in]{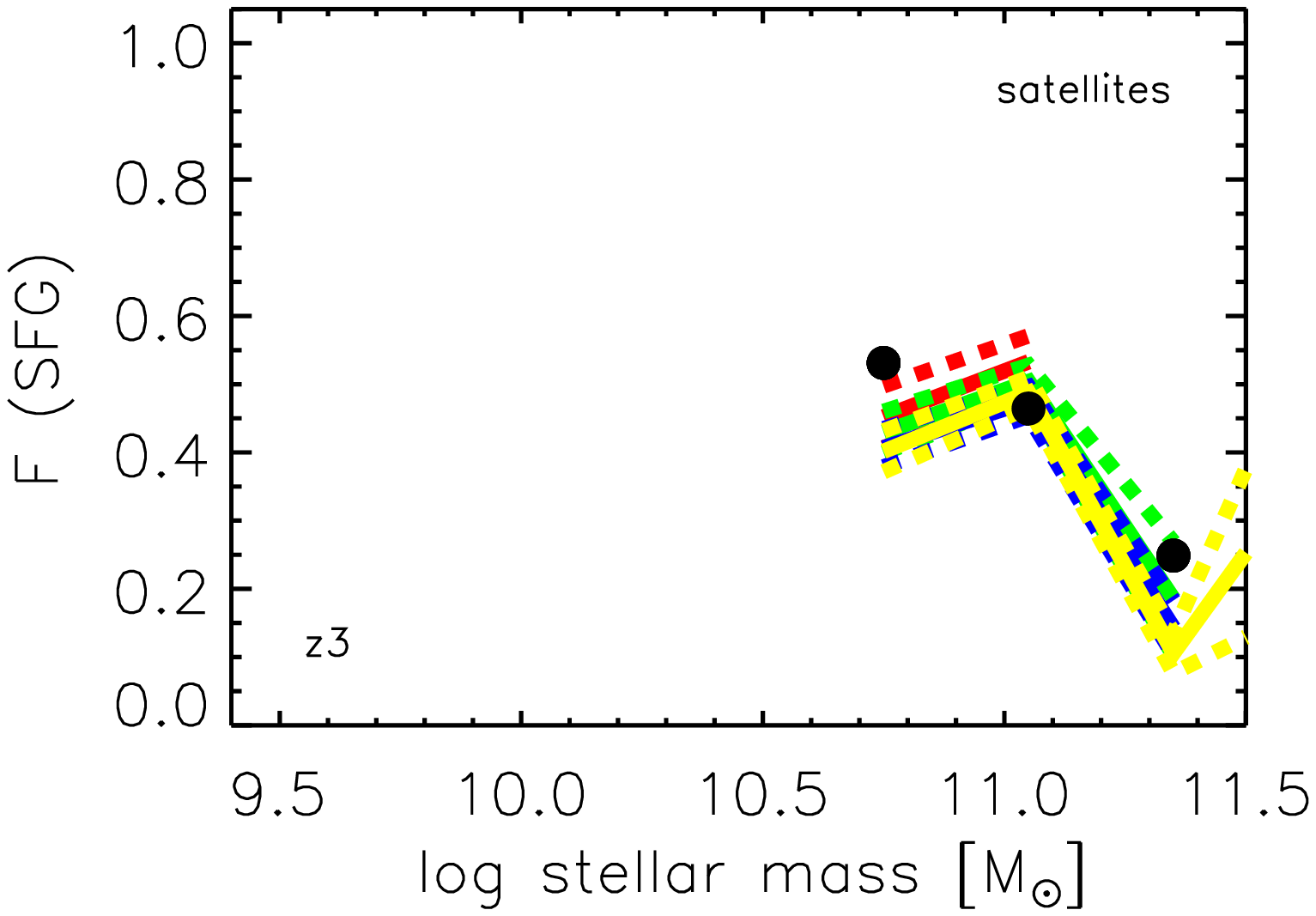}
\caption{${\rm F(SFG)}$ as a function of stellar mass. The three rows correspond to the three redshift bins. Field galaxies (filled circles) are compared with centrals (left) and satellites (right) in four halo mass bins.}
\label{fsf_mstar}
\end{figure}

To avoid issues related to redshift evolution of the environmental dependence, we only compare our results with similar observational studies in roughly the same redshift range. The existing results are quite mixed and it is difficult to do a direct and robust comparison due to different measures of environment (e.g., halo mass vs. density field) and different definitions of star-forming galaxies, in addition to other issues such as sample size and different methods of analysis. von der Linden et al. (2010) showed that the typical SFR of non-quiescent galaxies decreases by $\sim0.3$ dex towards the centre of clusters at $z<0.1$. Peng et al. (2010) found that the SSFR of blue galaxies is independent of environment (characterised by density quartiles). In Peng et al. (2012), a small difference in the SSFR as a function of stellar mass exists between central and satellite star-forming galaxies, which is $\sim0.2$ dex at low masses and decreases with increasing stellar mass. Wijesinghe et al.(2012) concluded that the SFR of star-forming galaxies is independent of the density (based on a 5th nearest neighbour metric) at fixed stellar mass. Haines et al. (2013) found the SSFR of star-forming cluster galaxies are systematically lower (by $\sim$28\%) than comparable galaxies in the field at $0.1<z<0.3$. In Darvish et al. (2016), the SFR and SSFR of star-forming galaxies are found to be independent of the density field (including the lowest redshift bin $z=[0.1, 0.5]$). Darvish et al. (2017) concluded that the median SFR for star-forming galaxies is $\sim0.3-0.4$ dex lower in satellites and centrals in clusters compared to field galaxies at $z\lesssim0.5$.

\section{Conclusions and discussions}

We take advantage of the GAMA survey to study the environmental dependence of the fraction of SFG and the properties of the MS at $z<0.3$. Thanks to the depth and quality of GAMA data, we are able to examine environmental dependence (characterised by host halo mass) of the MS using stellar mass selected samples without preselecting star-forming galaxies. Furthermore, we can separate galaxies in group environment into centrals and satellites and separately study any dependence on halo mass. Our main conclusions are:

\begin{itemize}
\item A clear bimodality exists in the SSFR distribution at fixed stellar mass which can be described by the combination of two Gaussians (the star-forming Gaussian and the quiescent Gaussian).

 \item Using the observed bimodality to define SFG, we check how the MS changes with environment. For centrals, the position of the MS is similar to field galaxies but the width of the MS is somewhat larger. No significant dependence on halo mass is found. For satellites, the position of the MS is almost always lower (by $\sim0.2$ dex) compared to field galaxies and the width is almost always larger. We do not find significant dependence of the position of the MS on halo mass.

\item The fraction of SFG ${\rm F(SFG)}$ is similar between centrals and field galaxies. We do not detect significant dependence on halo mass for the centrals. However, for satellites, ${\rm F(SFG)}$ decreases with increasing halo mass and this dependence on halo mass strengthens significantly towards lower redshift.

\end{itemize}

\begin{acknowledgements}
LW thanks Ying-jie Peng for stimulating discussions on the galaxy MS. PN acknowledges the support of the Royal Society through the award of a University Research Fellowship, the European Research Council, through receipt of a Starting Grant (DEGAS-259586) and the support of the Science and Technology Facilities Council (ST/L00075X/1). GAMA is a joint European-Australasian project based around a spectroscopic campaign using the Anglo- Australian Telescope. The GAMA input catalogue is based on data taken from the Sloan Digital Sky Survey and the UKIRT Infrared Deep Sky Survey. Complementary imaging of the GAMA regions is being obtained by a number of in- dependent survey programmes including GALEX MIS, VST KiDS, VISTA VIKING, WISE, Herschel-ATLAS, GMRT and ASKAP providing UV to radio coverage. GAMA is funded by the STFC (UK), the ARC (Australia), the AAO, and the participating institutions. The GAMA website is http://www.gama-survey.org/.
\end{acknowledgements}

\newpage

\begin{appendix}

\section{Galaxy bimodality as described by the combination of two Gaussian distributions}

\begin{table}
\centering
\caption{Gaussian fits to the $\log$ SSFR distribution. The columns are galaxy type, the location and scale parameter of the star-forming Gaussian, $\mu_{\rm SF}$ (in unit of Gyr$^{-1}$) and $\sigma_{\rm SF}$ (in dex), the location and scale parameter of the quiescent Gaussian, $\mu_{\rm Q}$ (in unit of Gyr$^{-1}$) and $\sigma_{\rm Q}$ (in dex), the threshold $T$  (in unit of Gyr$^{-1}$) in $\log$ SSFR where the amplitudes of two Gaussians are equal to each other, and the number of galaxies.}\label{table:selection}
\begin{tabular}{lllllll}
&&$z1$ &= [0.01, 0.1]&&&\\
Type &  $\mu_{\rm SF}$   & $\sigma_{\rm SF}$ &   $\mu_{\rm Q}$ & $\sigma_{\rm Q}$& T & N\\
field      &         -10.0&     0.4&     -11.5&     0.7&     -10.7  & 3981\\
central     &     -10.3&     0.5&     -12.1&     0.5&     -11.1    & 762\\
satellite     &       -10.2&     0.5&     -11.8&     0.5&     -11.0  & 2146\\
&&$z2$& = [0.1, 0.2]&& &\\
field      &        -10.3&     0.4&     -11.6&     0.4&     -10.9  & 8500\\
central     &        -10.4&     0.5&     -11.9&     0.4&     -11.1  & 2232\\
satellite     &     -10.6&     0.5&     -11.7&     0.4&     -11.1& 4250 \\
&&$z3$ &= [0.2, 0.3]&&&\\
field      &     -10.5&     0.4&     -11.5&     0.3&     -11.0 & 7037\\
central     &       -10.8&     0.6&     -11.7&     0.3&     -11.2& 1972 \\
satellite     &    -10.8&     0.5&     -11.6&     0.3&     -11.2  &  2806\\
\end{tabular}
\end{table}

\end{appendix}
\end{document}